\documentclass[aps,prx,twocolumn,amsmath,amssymb,nofootinbib,superscriptaddress,floatfix,reprint,longbibliography]{revtex4-1}

\usepackage[dvips]{graphicx}
\usepackage{latexsym}
\usepackage{amsmath}
\usepackage{amsfonts}
\usepackage{amssymb}
\usepackage{bm}
\usepackage{color}
\usepackage{txfonts}
\usepackage{float}
\usepackage{braket}
\usepackage{url}
\usepackage{CJKutf8} 
\usepackage{svg}
\usepackage{multirow}
\usepackage{comment}
\usepackage[colorlinks=true, allcolors=blue]{hyperref}
\usepackage{ulem}
\usepackage{cleveref}

\newcommand{\C}[1]{{\textcolor{red}{#1}}}
\newcommand{\B}[1]{{\textcolor{blue}{#1}}}

\crefname{Supplement}{Supp.\,}{Supps.\,}

\creflabelformat{Supplement}{[#2#1#3]}

\begin{document}

\title{Revealing Hidden Unconventional Pairing through Nonreciprocal Transport}

\author{Wen-Bo Dai}
\affiliation{International Center for Quantum Materials, School of Physics, Peking University, Beijing 100871, China}
\affiliation{Beijing Academy of Quantum Information Sciences, Beijing 100193, China}
\affiliation{Department of Physics, Hong Kong University of Science and Technology, Clear Water Bay, Hong Kong, China}
\author{Ming Gong}
\affiliation{International Center for Quantum Materials, School of Physics, Peking University, Beijing 100871, China}
\affiliation{Department of Physics, The University of Tokyo, 7-3-1 Hongo, Tokyo 113-0033, Japan}
\author{Xianxin Wu}
\affiliation{CAS Key Laboratory of Theoretical Physics, Institute of Theoretical Physics,
Chinese Academy of Sciences, Beijing 100190, China}
\author{Chui-Zhen Chen}
\email{czchen@suda.edu.cn}
\affiliation{School of Physical Science and Technology, Soochow University, Suzhou 215006, China}
\affiliation{Institute for Advanced Study, Soochow University, Suzhou 215006, China}
\author{X. C. Xie}
\email{xcxie@pku.edu.cn}
\affiliation{International Center for Quantum Materials, School of Physics, Peking University, Beijing 100871, China}
\affiliation{Interdisciplinary Center for Theoretical Physics and Information Sciences, Fudan University, Shanghai 200433, China}
\affiliation{Hefei National Laboratory, Hefei 230088, China}
\date{\today }

\begin{abstract}
Identifying the pairing symmetry of Cooper pairs is a fundamental step toward understanding the microscopic mechanisms of unconventional superconductors. However, experimental identification remains a formidable challenge, particularly when unconventional pairing is obscured by a dominant $s$-wave component that masks its spectroscopic signatures.
Here, we develop a symmetry-resolved framework to identify superconducting pairing symmetry through nonreciprocal conductance upon exchanging source and detector terminals in multiterminal devices.
We show that nonreciprocal transport arises from symmetry-breaking components of the superconducting order parameter and exhibits a characteristic angular dependence that encodes the momentum-space structure of the pairing gap. In particular, time-reversal-breaking singlet pairing induces nonreciprocal charge transport, while spin-triplet pairing generates nonreciprocal spin responses, providing distinct transport fingerprints of the underlying order. We demonstrate this mechanism using representative models of iron-based and noncentrosymmetric superconductors and outline experimental protocols for multiterminal measurements. Our results advance the theoretical understanding of nonreciprocal transport in superconductors, and establish it as a symmetry-selective probe for identifying hidden unconventional pairing in a wide range of superconducting materials.
\end{abstract}
\maketitle

\section{Introduction}
The pairing symmetry of Cooper pairs is a defining property of a superconductor, 
 encoding the underlying microscopic pairing mechanism~\cite{BCS,mineev1999introduction}.
Identifying this symmetry is therefore of fundamental importance in condensed matter physics  \cite{RMPstriplet,symmetry_cuprate_RMP,RMPSZX}, with direct implications for resolving long-standing questions, such as the nature of high-temperature superconductivity \cite{HTcSC, SZXcuSC, symmetry_cuprate_RMP,RMPSZX,Cao2018,d+idmoire,Oh2021,Lothman2022} and advancing the search for topological superconductors \cite{RMPstriplet,Alicea_2012,Wilczek2009Majorana,TI2TSCPRB,FuKanePRL,QAH2TSCPRB,TI&TSCRMP}.
Despite its importance, determining the pairing symmetry remains a formidable experimental challenge, particularly in superconductors  that  break inversion  \cite{Gorkov2001,Bauer2004,Sigrist2004,Kimura2005,Nagaosa2009,KTLaw2014,Nagaosa2018,Ando2020,Hamill2021} or time-reversal symmetry \cite{CongjunWu,XiaoyuZhu,Khodas,Fernandes2022,Kheirkhah,Grinenko2020,Mallik}.
In such materials, an unconventional pairing state often coexists with a conventional s-wave component \cite{Mallik,PRL2016ZXS}.
This mixing obscures hallmark signatures of the unconventional pairing, such as unconventional gap structures \cite{RMPSZX,gaoARPESNC2024} or phase-sensitive features \cite{symmetry_cuprate_RMP}, especially when the s-wave component is dominant \cite{PNAS2012ZXS,PRL2013Razzoli,PRB1995Devereaux,PPRX2014WHH}.
Consequently, widely used spectroscopic probes, such as angle-resolved photoemission spectroscopy (ARPES) and scanning tunneling microscopy/spectroscopy (STM/STS), which are primarily sensitive to gap-related spectral amplitudes,
may fail to resolve the hidden unconventional component.		
\begin{figure*}[bht]
    \centering
    \includegraphics[width=5.5in]{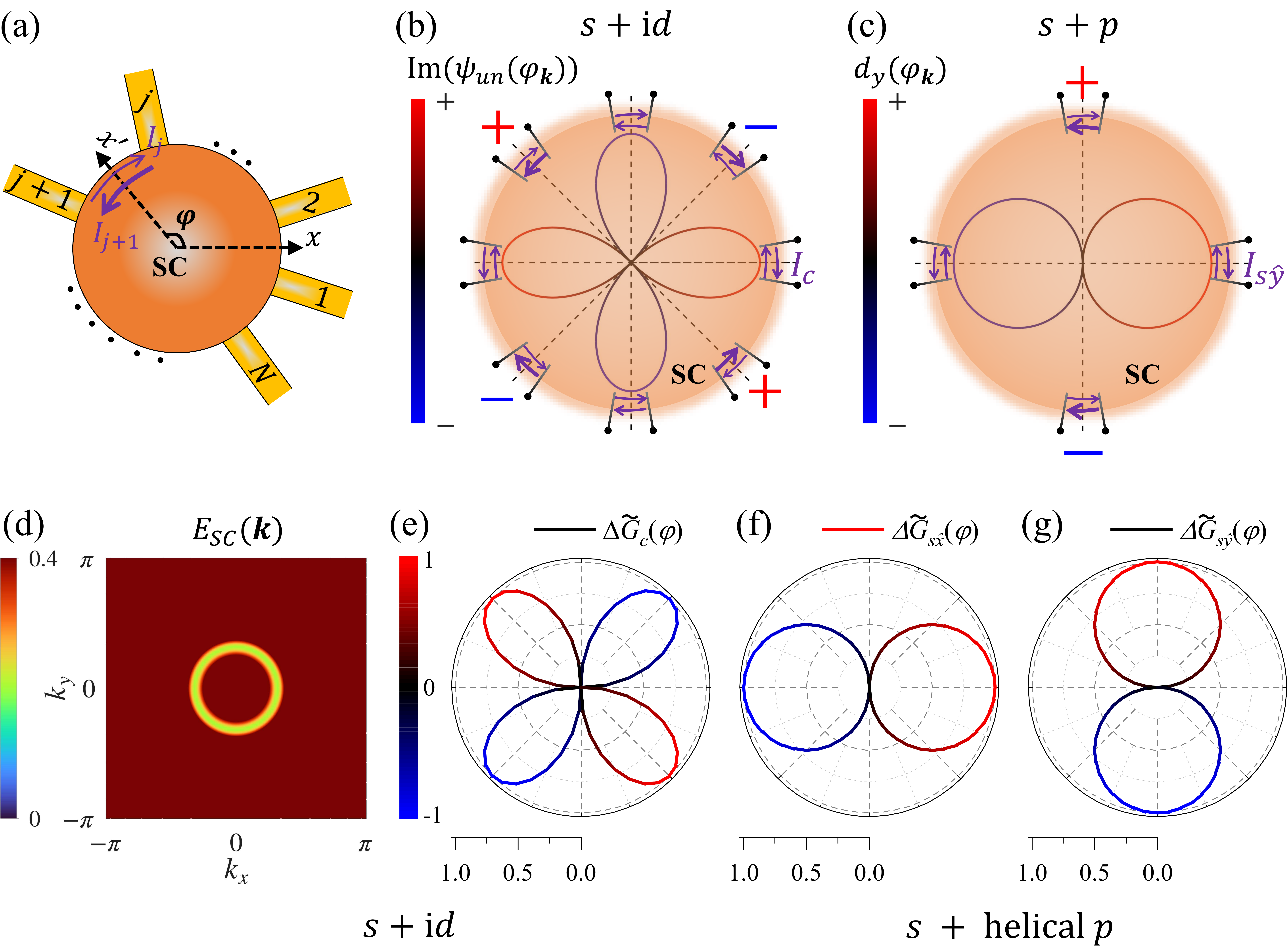}
    \caption{(Color online). 
Nonreciprocal detection scheme.
(a) Schematic of nonreciprocal transport in a superconducting system (orange) connected to $N$ terminals (gold). 
SC represents the superconducting region.
A voltage bias applied to terminal $j$ generates a charge or spin current $I_{j+1}$ into terminal $j+1$, while the reverse process induces $I_j$. 
The direction $x'$ indicates the local interface normal, associated with the normal angle $\varphi$. 
(b–c) Angular distributions of the pairing components and associated nonreciprocal transport. Panels (b) and (c) display the time-reversal-breaking component $\mathrm{Im}[\psi_{\rm un}(\varphi_{\boldsymbol{k}})]$ for an $s+\mathrm{i}d$ state and the $\boldsymbol{d}$-vector component $d_y(\varphi_{\boldsymbol{k}})$ for an $s+p$ state, respectively. Red (blue) colors denote positive (negative) values of the pairing components. Purple arrows indicate the forward and backward transport of $I_c$ and $I_{s\hat{n}}$, whose differences define the charge and spin nonreciprocity, respectively. The $\pm$ markers locate the angular positions of the positive and negative extrema.
(d) Colormap of the quasiparticle spectrum \(E_{\rm SC}(\mathbf{k})\) for an \(s+\mathrm{i}d\) superconductor in the \(s\)-wave-dominated regime. The nearly isotropic gap illustrates that the unconventional \(d\)-wave component can be hidden in conventional spectral measurements.
(e-g) Polar plots of  the normalized nonreciprocal conductances $\Delta\Tilde{G}(\varphi)\equiv\Delta G(\varphi)/\mathrm{max}(|\Delta G(\varphi)|)$ versus angle $\varphi$.
(e) $\Delta\Tilde{G}_{c}$ versus $\varphi$ for $s~+~\mathrm{i}d$-wave 
(d-e) consider $s~+~\mathrm{i}d$-wave with $\boldsymbol{d}=\boldsymbol{0}$ and $\psi_{un}=\mathrm{icos}(2\varphi_{\boldsymbol{k}})$.
(f) $\Delta\Tilde{G}_{s\hat{x}}$,
and (g) $\Delta\Tilde{G}_{s\hat{y}}$ versus $\varphi$ for s + helical p-wave with $\boldsymbol{d}=-\hat{x}\mathrm{sin}(\varphi_{\boldsymbol{k}})+\hat{y}\mathrm{cos}(\varphi_{\boldsymbol{k}})$ and $\psi_{un}=0$. 
A quadratic dispersion relation $\epsilon({\boldsymbol{k}})=Bk^2$ with $B=1$, $\mu=1$, $\Delta_s=0.2$ and $\lambda=0.02$ is used for (d-g).
\label{fignonre} }
    \end{figure*}

In this work, we demonstrate that nonreciprocal transport offers a symmetry-resolved strategy to unveil these hidden unconventional pairing channels.
Rooted in the fundamental symmetry principles of transport reciprocity—unifying the generalized Onsager framework~\cite{Onsager1931_1,Onsager1931_2,CasimirRMP1945,ButtikerPRL1986,Buttiker1988,datta1995} with recent advances in nonreciprocal phenomena~\cite{Tokura2018,NagaosaPRL2018,Ando2020,Lin2022,Trahms2023}—we establish a symmetry-enforced correspondence between superconducting order and nonreciprocal transport.
Operationally, nonreciprocity appears as a difference in linear-response conductance upon exchanging the source and detector terminals \cite{CasimirRMP1945,ButtikerPRL1986,Buttiker1988,datta1995} of a multiterminal device (see Fig.~\ref{fignonre}(a)).
We find that a pure $s$-wave  pairing, preserving all relevant symmetries, exhibits perfectly reciprocal transport.
In contrast, the emergence of unconventional pairing components that break the relevant symmetries generate characteristic nonreciprocal conductances. 
As a result, this transport nonreciprocity acts as a symmetry-enforced filter: the conventional s-wave component produces no intrinsic nonreciprocal signal, effectively removing the s-wave background, allowing the underlying unconventional order to be isolated. 
In particular, nonreciprocal charge and spin transport encode distinct symmetry sectors of the superconducting order parameter, corresponding respectively to time-reversal symmetry-breaking singlet components and triplet pairing channels as shown in Fig.~\ref{fignonre}(b-c).
Crucially, by probing symmetry rather than gap magnitude, these signatures remain robust against a dominant $s$-wave background, overcoming a central limitation of conventional spectroscopy.
This is verified by the angular correspondence shown in Fig.~\ref{fignonre}(e–g).
At the microscopic level, these symmetry-dependent fingerprints trace back to the lifting of symmetry constraints on scattering processes—a mechanism illustrated in Fig.~\ref{figscatt}(a,b) and summarized in Table~\ref{tab:tableexamples}.
We substantiate this proposal through numerical simulations for candidate materials\cite{Khodas,Gorkov2001}, and outline experimental protocols to realize such measurements in Fig.~\ref{figset}\cite{Kang2019NonlinearAHE,Bachmann2022DirectionalBallistic,Cherepanov2012,lupke2015scanning,Lupke2017DefectResistance,Baringhaus2014BallisticGraphene,Gerasimenko2019QuantumJamming,Kolmer2019}.
Our results establish nonreciprocal transport as a \B{symmetry-selective} route for identifying unconventional pairing in a wide range of superconducting materials, including iron-based\cite{Khodas,CongjunWu,KretzschmarPRL2013}, kagome\cite{OrtizKagomePRL2020,mielkeKagomeNature2022,fengKagomeNC2025}, moiré\cite{XCkPRL2018,ScheurerPRR2020,BalentsNP2020,CaoNature2021}
and noncentrosymmetric superconductors\cite{Bauer2004,FrigeriPRL2004,MukudaPRL2008,PRLYuanHQ}.

\section{Results}
\label{sec:Res}

\subsection{Generic model}
Fig.~\ref{fignonre}(a) illustrates a superconducting region connected to multiple terminals.
The superconducting region is modeled by a general effective Hamiltonian:
\begin{eqnarray}\label{eqH}
H_{SC} = H_s + \lambda V_{un},
\end{eqnarray}
which incorporates a mixture of conventional \(s\)-wave pairing \(H_s\) and an unconventional pairing component \(V_{\mathrm{un}}\) with strength \(\lambda\).
$V_{un}(\varphi_{\boldsymbol{k}})$ is  assumed to rely on the momentum angle $\varphi_{\boldsymbol{k}}$ only.
Accordingly, the superconducting pairing gap matrix, $\hat{\Delta}$ takes the form
\begin{equation}
\hat{\Delta}(\varphi_{\boldsymbol{k}}) = \mathrm{i} \Big[ \Delta_s + \lambda \big(\psi_{\rm un}(\varphi_{\boldsymbol{k}}) + \boldsymbol{d}(\varphi_{\boldsymbol{k}})\cdot \boldsymbol{\sigma} \big) \Big] \sigma_y.
\label{eq:gap_matrix}
\end{equation}
where $\Delta_s$ denotes the conventional $s$-wave pairing gap, taken to be real for simplicity.
Here, $\psi_{un}(\varphi_{\boldsymbol{k}})$ denotes the unconventional spin-singlet pairing amplitude, while $d_n(\varphi_{\boldsymbol{k}})$ represents the $\hat{n}$ component of the spin-triplet $\boldsymbol{d}$-vector.
In the limit \(\lambda = 0\), the system reduces to a pure \(s\)-wave superconductor.

To elucidate the nonreciprocity, we compare the current response between two adjacent terminals labeled \( j \) and \( j+1 \).  
As shown in Fig.~\ref{fignonre}(a),
 applying a voltage to terminal \( j \)  induces a current  \( I_{j+1} \) in terminal  \( j+1 \) as  illustrated by the thick purple arrow.
Reversing the configuration,  a voltage in terminal \( j +1 \) yields a current   \( I_j \) in terminal  \( j \), represented by the thin purple arrow in the opposite direction.  
Such a finite difference between the two reversed configurations, \( I_j \neq I_{j+1} \), under an identical voltage bias, serves as a definitive signature of nonreciprocal transport.

\subsection{Symmetry-correspondence}
This nonreciprocal transport directly encodes the symmetry structure of unconventional pairing. 
In a conventional $s$-wave superconductor without unconventional pairing($\lambda=0$), transport nonreciprocity is symmetry-forbidden: time-reversal symmetry forbids nonreciprocal charge transport, while spin-rotation symmetry forbids nonreciprocal spin transport. 
In contrast, when $\lambda \neq 0$, an unconventional pairing potential $V_{\mathrm{un}}$ can selectively break these symmetries, thereby inducing measurable nonreciprocal transport responses.

For a  spin-singlet unconventional pairing which breaks time-reversal $\mathcal{T}$ symmetry, the nonreciprocal transport of charge current $I_{c}$ emerges.
It is governed by the $\mathcal{T}$-breaking component $\mathrm{Im}[\psi_{\rm un}(\varphi_{\boldsymbol{k}})]$, which introduces a characteristic $d$-wave angular anisotropy. 
As exemplified by the $s+\mathrm{i}d$ state in Fig.~\ref{fignonre}(b), the nonreciprocal response of $I_c$ vanishes at the extrema of $\mathrm{Im}(\psi_{\rm un})$, where the $\mathcal{T}$-breaking component $\mathrm{Im}(\psi_{\rm un})$ is invariant under mirror reflection perpendicular to the transport direction, and is maximized near its nodal directions, where this mirror symmetry of $\mathrm{Im}(\psi_{\rm un})$ is maximally broken.
Consequently, its angular dependence directly reflects that of $\mathrm{Im}[\psi_{\rm un}(\varphi_{\boldsymbol{k}})]$, with a $\pi/4$ rotational offset set by the $d$-wave angular anisotropy, as illustrated in Fig.~\ref{fignonre}(b).

Similarly, the nonreciprocal spin transport emerges in a spin-triplet unconventional pairing, which breaks spin-rotation symmetry.
As exemplified by the $s+p$ state in Fig.~\ref{fignonre}(c), the $d_n(\varphi_{\boldsymbol{k}})$ component of the $\boldsymbol{d}$-vector breaks spin-rotation symmetry $s_{n_\perp}$, where $n_\perp$ denotes the direction perpendicular to $\hat{n}$. 
Consequently, $d_n$ governs the nonreciprocal transport of spin current $I_{s\hat{n}}$, whose angular dependence tracks $d_n(\varphi_{\boldsymbol{k}})$, with a $\pi/2$ rotational offset set by the $p$-wave anisotropy.
Here, $I_{s\hat{n}}$ represents the current of the spin component along the $\hat{n}$ direction.

Notably, the correspondence established above follows solely from symmetry considerations of the pairing potential and is independent of model details.
The resulting nonreciprocal response is therefore symmetry-selective, which arises exclusively from a $\mathcal{T}$ or $s_{n_{\perp}}$ symmetry-breaking pairing component, rather than from geometrical asymmetry\cite{supp}.
\begin{figure}[bht]
    \centering
    \includegraphics[width=3.4in]{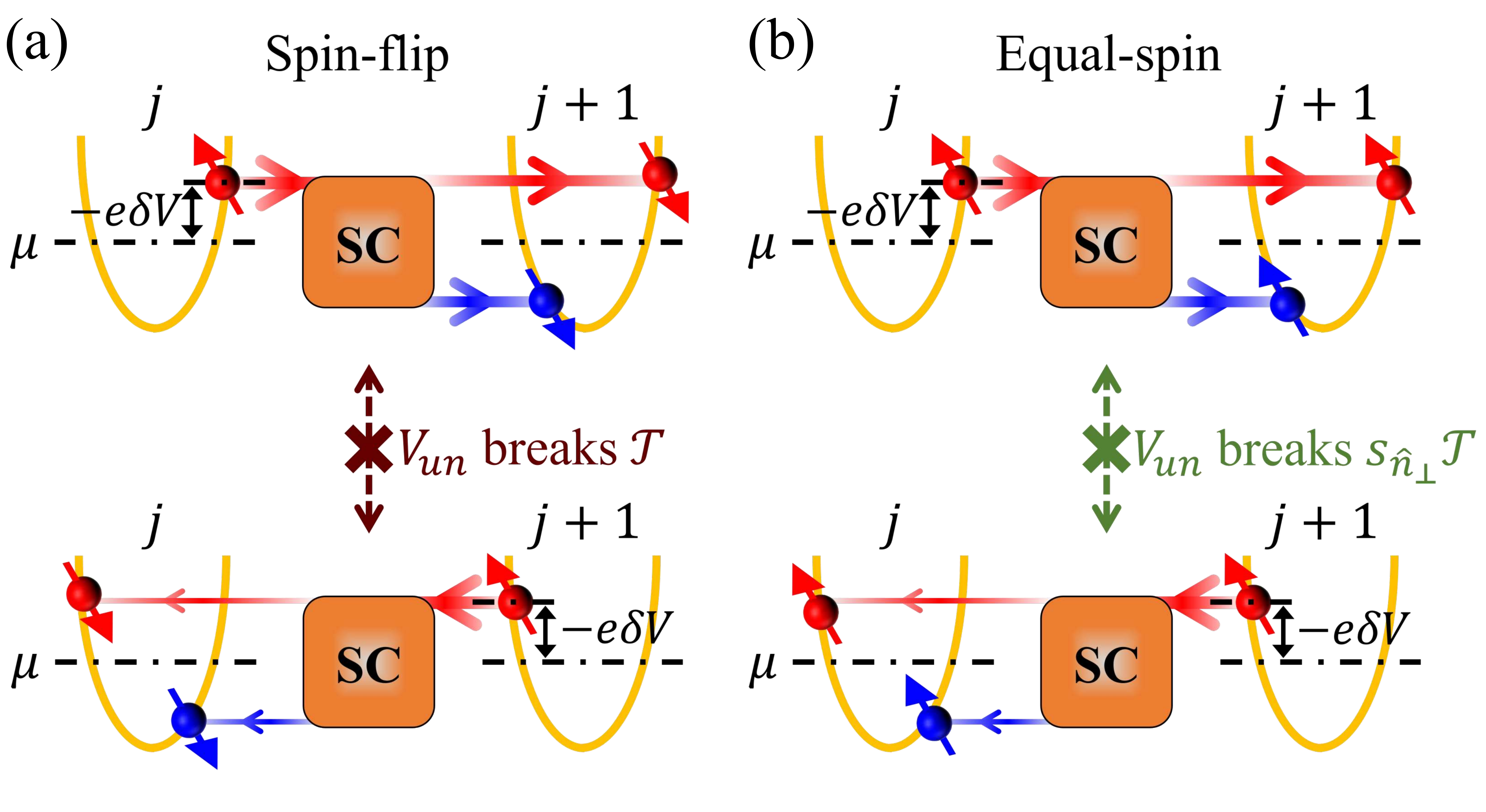}
    \caption{Microscopic origin of pairing-induced nonreciprocal transport.
    (a) Illustration of spin-flip scattering nonreciprocity, induced by time-reversal $\mathcal{T}$-breaking pairing \( V_{\mathrm{un}} \).
    Here SC represents the superconducting region (orange) and $\delta V$ denotes a small voltage bias.
    Top: The spin-flip scattering process from terminal $j$ to $j+1$.
    A spin-up electron from lead \( j \) is scattered into lead \( j+1 \) as a spin-down electron or a spin-down hole. 
    Bottom: The reverse process from terminal \( j+1 \) to \( j \). Line widths indicate injection/transmission probabilities.
    The red ball represents electron and the upward (downward) arrow represents spin-up (spin-down).
    (b) Similar to (a), but for the nonreciprocity of spin-preserving scattering induced by the spin–time-reversal symmetry \( s_{\hat{n}_\perp}\mathcal{T} \)-breaking $V_{un}$.
    \label{figscatt} }
    \end{figure}
\begin{table*}[bht]
\caption{\label{tab:tableexamples}
Correspondence between types of nonreciprocal conductances and representative unconventional pairings $V_{\mathrm{un}}$.
$\checkmark$ means the symmetry is preserved.
$\times$ means that a symmetry is broken or a nonreciprocal scattering channel is forbidden.
$\circ$ means that the nonreciprocal scattering process is allowed.}

\setlength{\tabcolsep}{4pt}      
\renewcommand{\arraystretch}{0.85} 
\small                           

\begin{ruledtabular}

\begin{tabular}{cccccc}

\multirow{2}{*}{Spin type of $V_{un}$}
&
\multirow{2}{*}{$\mathcal{T}$}
&
\multicolumn{2}{c}{Nonreciprocal Scattering}
&
\multirow{2}{*}{Examples}
&
\multirow{2}{*}{Nonreciprocal Conductances}
\\

&
&
spin-flip
&
equal-spin
&
&
\\
\hline

Spin-singlet
&
$\times$
&
$\circ$
&
$\circ$
&
$s+\mathrm{i}d,\ s+\text{chiral }d$
&
$\Delta G_{c}\propto
\mathrm{Im}(\psi_{\mathrm{un}}(\varphi+\frac{\pi}{4}))
$

\\[12pt]

\multirow{2}{*}{Spin-triplet}
&
\multirow{2}{*}{$\times$}
&
\multirow{2}{*}{$\circ$}
&
\multirow{2}{*}{$\circ$}
&
\multirow{2}{*}{$s+\text{chiral }p$}
&
\multirow{2}{*}{
$
\begin{array}{c}
\Delta G_{\uparrow_{\hat n},\uparrow_{\hat n}}
\propto
\mathrm{Re}(d_n(\varphi+\frac{\pi}{2}))
\\[-2pt]
\Delta G_{\downarrow_{\hat n},\uparrow_{\hat n}}
\propto
\mathrm{Im}(d_n(\varphi+\frac{\pi}{2}))
\end{array}
$
}
\\[6pt]

&
&
&
&
&
\\[-2pt]

&
$\checkmark$
&
$\times$
&
$\circ$
&
$s+p$
&
$\Delta G_{s\hat n}
\propto
d_n(\varphi+\frac{\pi}{2})
$

\end{tabular}

\end{ruledtabular}
\end{table*}

\subsection{Nonreciprocal conductances $\&$ minimal illustration}
These nonreciprocal transport phenomena are quantified by the nonreciprocal charge and spin conductances between adjacent terminals, defined as
\begin{equation}\label{eq:DGcs}
\Delta G_{c(s\hat{n})}(\varphi) = G_{c(s\hat{n}),\,j+1,j} - G_{c(s\hat{n}),\,j,j+1},
\end{equation}
where $G_{c(s\hat{n});\alpha,\beta}$ denotes the linear response of the charge (spin) current $I_{\alpha c(s\hat{n})}$ in terminal $\alpha$ to a voltage $V_{\beta}$ applied at terminal $\beta$.
Here, $\varphi$ is the spatial orientation angle, as illustrated in Fig.~\ref{fignonre}(a).

To demonstrate that the symmetry-based correspondence can be explicitly realized and quantitatively resolved in transport, we adopt a minimal Bardeen–Cooper–Schrieffer (BCS) Hamiltonian $H_{SC}=H_{\rm{BCS}}$ as an illustrative realization:
\begin{equation}
\begin{split}
H_{\rm{BCS}}(\boldsymbol{k}) &= [\epsilon(\boldsymbol{k}) - \mu]\, \sigma_0 \tau_z 
  + \left( \tau_+ \hat{\Delta}(\varphi_{\boldsymbol{k}}) + \mathrm{h.c.} \right) \\
\end{split}
\label{eq:BdGH}
\end{equation}
Notably, the minimal model is introduced to provide a concrete realization; the robustness of the correspondence against model details is demonstrated below using more realistic Hamiltonians.
Here $\epsilon (\boldsymbol{k})$ is the kinetic energy and $\mu$ is the chemical potential.
Pauli matrices $\tau_{x,y,z}$ ($\sigma_{x,y,z}$) act on particle-hole (spin) space and $\tau_{+}=\frac{1}{2}(\tau_x +i\tau_y)$.
The pairing matrix $\hat{\Delta}(\varphi_{\boldsymbol{k}})$ is defined as in Eq.~\ref{eq:gap_matrix}.

For an $s+\mathrm{i}d$ pairing, $\boldsymbol{d} = 0$ and $\psi_{\text{un}}(\varphi_{\boldsymbol{k}}) = \mathrm{i} \cos(2\varphi_{\boldsymbol{k}})$.
In the regime where the $d$-wave component is dominated by the $s$-wave gap (i.e., $\lambda \ll 1$), characteristic features of the nodal $d$-wave state, such as gap nodes, Fermi arcs, and the twofold rotational symmetry of the order parameter (fourfold in the gap magnitude)\cite{symmetry_cuprate_RMP,RMPSZX,supp}, are obscured by the isotropic $s$-wave gap, as illustrated by the energy dispersion in Fig.~\ref{fignonre}(d). 
In contrast, the nonreciprocal-conductance \B{scheme} developed here remains sensitive to the underlying $\mathcal{T}$-breaking structure.
Consistent with the symmetry analysis illustrated by Fig.~\ref{fignonre}(b), the simulated nonreciprocal charge conductance $\Delta G_c(\varphi)$ in Fig.\ref{fignonre}(e) exhibits a petal-shaped angular pattern, proportional to the angular profile of $\mathrm{Im}(\psi_{\text{un}}(\varphi_{\boldsymbol{k}}))$ rotated by $\pi/4$.

For an $s+p$ pairing, we consider an \(s\) + helical \(p\)-wave pairing \cite{Matano2016}, in which the $p$-wave pairing is strongly masked by a dominant $s$-wave gap (i.e., $\lambda \ll 1$).
In this case, the unconventional pairing is given by \(\boldsymbol{d} = -\hat{x} \sin(\varphi_{\boldsymbol{k}}) + \hat{y} \cos(\varphi_{\boldsymbol{k}})\) and \(\psi_{\text{un}} = 0\).
Fig.~\ref{fignonre}(f,g) demonstrate that the simulated nonreciprocal spin conductances $\Delta G_{s\hat{x}}(\varphi)$ and $\Delta G_{s\hat{y}}(\varphi)$ are proportional to the angular profiles of $d_x(\varphi_{\boldsymbol{k}})$ and $d_y(\varphi_{\boldsymbol{k}})$, respectively, each rotated by $\pi/2$, fully consistent with the correspondence illustrated in Fig.~\ref{fignonre}(c).

\subsection{Microscopic mechanisms}
The macroscopic transport asymmetry is a direct consequence of nonreciprocity at the microscopic level, originating from fundamental scattering processes within the superconductor.
To make this connection explicit, we decompose the nonreciprocal charge and spin conductances into spin-dependent contributions as
\begin{eqnarray} 
\Delta G_{c}(\varphi)
&=&
\sum_{\sigma,\sigma'}\Delta G_{\sigma,\sigma'} ,
\label{eq:DGc} \\
\Delta G_{s\hat{n}}(\varphi)
&=&
\frac{\hbar}{2e}
\sum_{\sigma'}
\Big(
\Delta G_{\uparrow_{\hat{n}},\sigma'}
-
\Delta G_{\downarrow_{\hat{n}},\sigma'}
\Big) .
\label{eq:DGs}
\end{eqnarray}
Here $\sigma,\sigma' \in \{\uparrow_{\hat{n}},\downarrow_{\hat{n}}\}$ label spin along the quantization axis $\hat{n}$ and $\Delta G_{\sigma,\sigma'} = G_{j+1\sigma,j\sigma'}-G_{j\sigma,j+1\sigma'}$ denotes the nonreciprocity of the spin-dependent conductance $G_{\alpha\sigma,\beta\sigma'}=\partial I_{\alpha\sigma}/\partial V_{\beta\sigma'}$.
$I_{\alpha\sigma}$ is the spin-dependent current in lead $\alpha$ and $V_{\beta\sigma'}$ is the spin-dependent bias voltage applied to lead $\beta$~\cite{Sun_2009,SunPRB2011}.
Within the Landauer–Büttiker formalism~\cite{Fisher-LeePRL,Wingreen1992,Wingreen1993}, each $\Delta G_{\sigma,\sigma'}$ directly reflects the asymmetry between forward and backward spin-dependent scattering processes,
\begin{equation}\label{eqspindepG}
\begin{aligned}
\Delta G_{\sigma,\sigma'}
&=
\frac{e^2}{h}
\Big[
\big(
T^{N}_{j+1\sigma,j\sigma'}
-
T^{N}_{j\sigma,j+1\sigma'}
\big)
\\
&\qquad
-
\big(
T^{A}_{j+1\sigma,j\sigma'}
-
T^{A}_{j\sigma,j+1\sigma'}
\big)
\Big].
\end{aligned}
\end{equation}
Here, $T^{A(N)}_{\alpha\sigma,\beta\sigma'}$ denotes the Andreev (normal) transmission probability from lead $\beta$ with spin $\sigma'$ to lead $\alpha$ with spin $\sigma$.
As depicted in Fig.~\ref{figscatt}(a) and (b), these processes can be broadly categorized into two distinct, spin-resolved channels: spin-flip scattering ($\sigma=\downarrow_{\hat{n}},~\sigma'=\uparrow_{\hat{n}}$) and equal-spin scattering ($\sigma=\uparrow_{\hat{n}},~\sigma'=\uparrow_{\hat{n}}$) respectively.

For a pure $s$-wave superconductor, transport reciprocity is protected in both spin-resolved scattering channels.
Time-reversal symmetry $\mathcal{T}$ enforces reciprocity in the spin-flip scattering channel,
$T^{N(A)}_{\alpha\downarrow_{\hat{n}},\beta\uparrow_{\hat{n}}} = T^{N(A)}_{\beta\downarrow_{\hat{n}},\alpha\uparrow_{\hat{n}}}$
by reversing both the scattering path and the spin orientations.
This immediately yields the spin-flip nonreciprocal conductance vanishes, $\Delta G_{\downarrow_{\hat{n}},\uparrow_{\hat{n}}}=0$.
Similarly, reciprocity in the equal-spin channel is protected by the combined spin–time symmetry $s_{\hat{n}_\perp}\mathcal{T}$,
which enforces $\Delta G_{\uparrow_{\hat{n}},\uparrow_{\hat{n}}}=0$.
Therefore, in the absence of unconventional pairing ($\lambda=0$), all spin-dependent nonreciprocal conductances $\Delta G_{\sigma,\sigma'}$ vanish.

By contrast, when $\lambda\neq 0$, an unconventional pairing $V_{\rm un}$ can selectively break
$\mathcal{T}$ and $s_{\hat{n}_\perp}\mathcal{T}$ symmetries, thereby generating
measurable nonreciprocity in the spin-flip and equal-spin scattering channels,
as illustrated in Fig.~\ref{figscatt}(a) and (b).
As a result, different components of the unconventional pairing contribute to distinct $\Delta G_{\sigma,\sigma'}$ via different spin-scattering channels, leading to finite nonreciprocal charge and spin responses according to Eqs.~\eqref{eq:DGc} and \eqref{eq:DGs}.

The correspondence following this microscopic decomposition is summarized in Table~\ref{tab:tableexamples},  which has been verified in supplementary materials numerically~\cite{supp}.
The first row concerns spin-singlet unconventional pairings.
The $\mathcal{T}$-breaking component $\mathrm{Im}\!\left[\psi_{\rm un}(\varphi)\right]$ breaks both $\mathcal{T}$ and $s_{\hat{n}_\perp}\mathcal{T}$ symmetries, giving rise to nonreciprocity in both spin-flip and equal-spin channels and hence a finite $\Delta G_c$.
For even orbital angular momentum $l$, the $l$-fold rotational symmetry further constrains the response to the form listed in the first row of Table~\ref{tab:tableexamples},
\begin{equation}
\Delta G_{c}(\varphi)\propto \mathrm{Im}\left[\psi_{\rm un}\left(\varphi+\frac{\pi}{2l}\right)\right].
\label{eqDGevenl}
\end{equation}

The second to third rows describe spin-triplet pairings, in which different $\boldsymbol{d}$-vector components selectively activate distinct spin-scattering channels: $\mathrm{Im}(d_{n})$ breaks $\mathcal{T}$ symmetry and generates $\Delta G_{s\hat{n}}$ via spin-flip scattering, while $\mathrm{Re}(d_{n})$ breaks $s_{\hat{n}_\perp}\mathcal{T}$ symmetry and contributes through equal-spin scattering.
For odd $l$, rotational symmetry constrains their contributions as
\begin{eqnarray}\label{eqDGoddl1}
	\left\{
	\begin{array}{lll}
        \Delta G_{\uparrow_{\hat{n}},\uparrow_{\hat{n}}}(\varphi) &\propto&\mathrm{Re}(d_{n}(\varphi+\frac{\pi}{2l}))\\
        \Delta G_{\downarrow_{\hat{n}},\uparrow_{\hat{n}}}(\varphi)&\propto&\mathrm{Im}(d_{n}(\varphi+\frac{\pi}{2l}))\\
    \end{array}
	\right.
\end{eqnarray}
as summarized in the second row of Table~\ref{tab:tableexamples}.

\begin{figure*}[bht]
    \centering
    \includegraphics[width=6.8in]{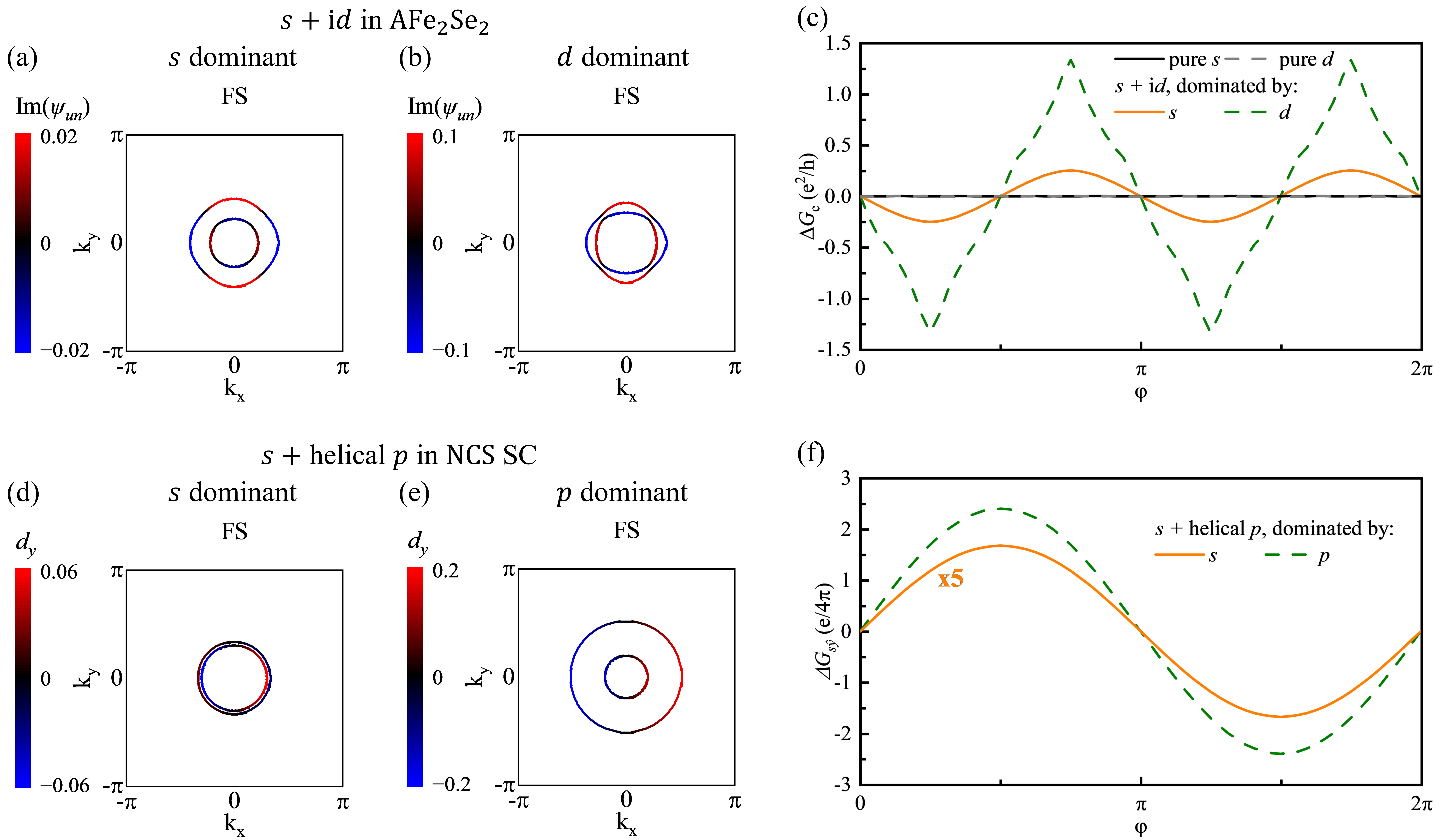}
    \caption{Pairing symmetry and nonreciprocal-transport fingerprints in candidate superconductors or material-specific nonreciprocal fingerprints of hidden unconventional pairing.
    (a-c) Pairing and nonreciprocal charge conductance in AFe$_2$Se$_2$
    (a,b) Plots of the imaginary part $\mathrm{Im}[\psi(\boldsymbol{k})]$ on the Fermi surfaces.
    (c) Plots of the nonreciprocal charge conductance $\Delta G_c(\varphi)$ versus angle $\varphi$.
    (d-f) Pairing and nonreciprocal spin conductance in noncentrosymmetric superconductors (NCS SC).
    (d,e) Plots of the $d_y$ on the Fermi surfaces.
    (f) Plots of the nonreciprocal spin conductance $\Delta G_{s\hat{y}}(\varphi)$ versus angle $\varphi$, where the curve for the $s$-wave-dominated case is amplified by a factor of 5.
    \label{figFeNonC} }
\end{figure*}
Accordingly, $\Delta G_{s\hat{n}}(\varphi)$ provides transport fingerprints of spin-triplet pairing channels and exhibits an $l$-fold rotational symmetry that reflects the orbital angular momentum of the spin-triplet unconventional pairing.
When $d_{n}$ is $\mathcal{T}$-invariant, the spin-flip contribution $\Delta G_{\downarrow_{\hat{n}},\uparrow_{\hat{n}}}$ vanishes due to the absence of the $\mathrm{Im}(d_{\hat{n}})$ component, leaving only the equal-spin channel active.
The resulting nonreciprocal spin conductance therefore provides a direct measure of the angular structure of the $\boldsymbol{d}$ vector,
\begin{align}\label{eqDGoddl2}
\Delta G_{s\hat{n}}(\varphi) \propto d_{n}\left(\varphi+\tfrac{\pi}{2l}\right),
\end{align}
as listed in the third row of Table~\ref{tab:tableexamples}.

In this way, $\Delta G_c$ selectively probes the angular structure of $\mathcal{T}$-breaking spin-singlet states, while $\Delta G_{s\hat{n}}$ resolves both the angular dependence and the spin-pairing channel of triplet states.
Moreover, $\Delta G_{\sigma,\sigma'}$ provides a phase-sensitive probe of triplet pairing.

\subsection{Experimental implications}
To demonstrate the experimental relevance of our framework, we apply it to pairing states proposed in candidate materials.

The $s+\mathrm{i}d$ state in the first row \B{of Table~\ref{tab:tableexamples}}  has been extensively studied in iron-based superconductors \cite{RMP2011,CongjunWu,PlattPRB2012,Khodas,Fernandes2022,Kheirkhah,Grinenko2020,Mallik}, and has been suggested as a possible pairing state in kagome systems\cite{OrtizKagomePRL2020,mielkeKagomeNature2022,fengKagomeNC2025}.
Experimental evidence suggests that the $d$-wave component can be obscured by a $s$-wave gap \cite{PPRX2014WHH}, rendering it effectively invisible to conventional probes. 
To demonstrate how nonreciprocal transport overcomes this limitation, we consider AFe$_2$Se$_2$ (A=K, Rb, Cs) \cite{Khodas} as an illustration. 
The system has two orbitals with orthogonally oriented elliptical Fermi surfaces (FSs), where superconducting pairing is dictated by the competition between intra-band interaction (favoring $d$-wave pairing) and inter-band interaction (favoring $s$-wave pairing).
This interplay is tuned by $\kappa$, the ratio of the FSs' hybridization strength to the FSs' eccentricity.
As $\kappa$ increases, the system undergoes a phase transition from a pure $d$-wave state to an $s+\mathrm{i}d$ state, and finally to a pure $s$-wave state.

Since only spin-singlet channels exist, the pairing on the FSs can be captured by a scalar \( \psi \).
The real part of the projected order parameter $\psi$ corresponds to the $s$-wave component, while the imaginary part $\mathrm{Im}(\psi)$ corresponds to the $d$-wave component, as dictated by our gauge choice. 
In the $s+\mathrm{i}d$ mixed phase, the $d$-wave pattern of $\mathrm{Im}(\psi)$ is visualized for both $s$-dominated and $d$-dominated regimes
in \ref{figFeNonC}(a-b) respectively. 
The nonreciprocal conductance $\Delta G_c$ versus the spatial orientation angle $\varphi$ for different superconducting phases is obtained numerically and shown in Fig.~\ref{figFeNonC}(c).
Pure $s$-wave and $d$-wave states yield no nonreciprocal response, whereas the $s+\mathrm{i}d$ phase exhibits a finite signal in both $s$- and $d$-dominated regimes. 
The angular dependence of $\Delta G_c$ in the $s+\text{i}d$ phase follows the $d$-wave pattern of $\mathrm{Im}(\psi)$ in Fig.~\ref{figFeNonC}(a-b), rotated by $\pi/4$.
\begin{figure*}[bht]
    \centering
    \includegraphics[width=5in]{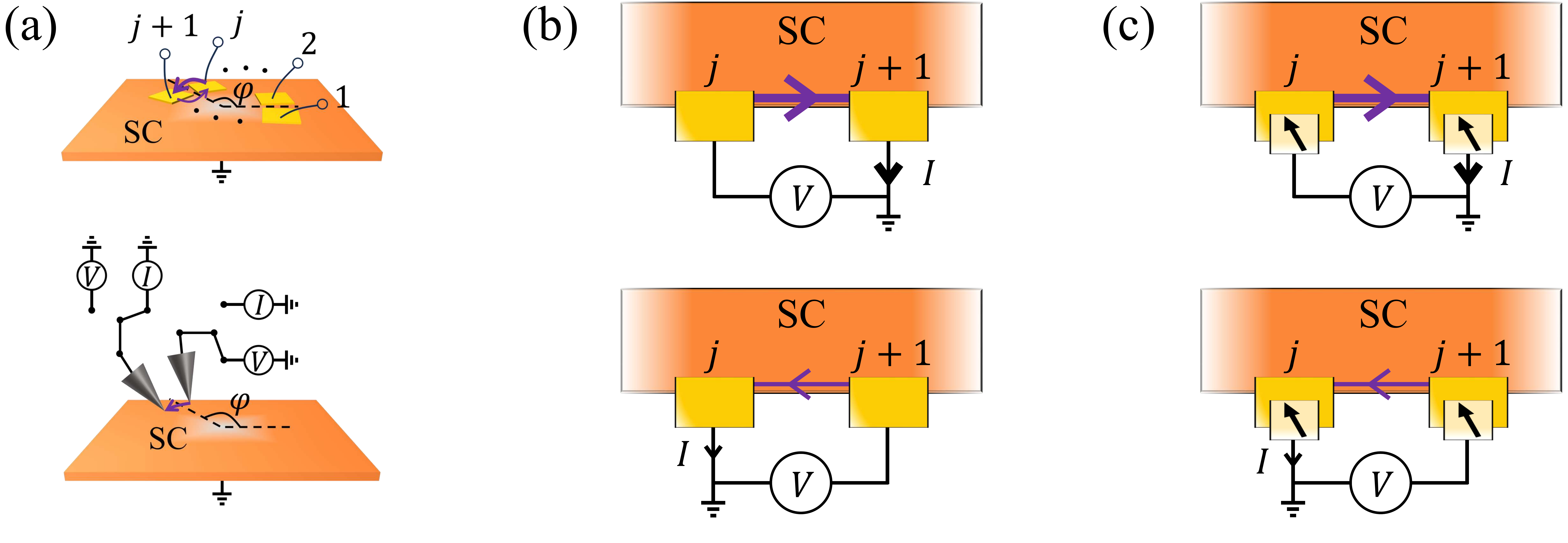}
    \caption{Experimental implementation of symmetry-resolved nonreciprocal transport or experimental protocols for measuring nonreciprocal conductance. 
    (a) Multi-tip STM configuration for nonlocal differential conductance. SC represents the superconducting region (orange). At angle $\varphi$, the source probe injects current and measures voltage ($V$), while the drain probe extracts and monitors current ($I$). Exchanging probe roles reverses the current path (purple arrow) to probe the nonreciprocal component.
    (b) 
    A pairwise measurement within a multiterminal
    measurement configuration for detecting nonreciprocal transport between leads $j$ and $j+1$. 
    The reverse-bias configuration is obtained by exchanging the source and detector terminals.    
        The nonreciprocal  current is obtained by exchanging source and drain terminals within the same multiterminal geometry.
        An asymmetry in the measured current under bias reversal indicates nonreciprocity.
    (c) Similar to (b), but with spin-polarized leads to probe spin-dependent nonreciprocal transport. 
        The black arrows indicate the spin polarization of each lead. 
    \label{figset} }
\end{figure*}

Another key scenario is the spin-triplet pairing admixed with an $s$-wave component, which is a widely studied in noncentrosymmetric superconductors (NCS SC)~\cite{Gorkov2001,Sigrist2004,Kimura2005,Nagaosa2009,Nagaosa2018,Ando2020,Hamill2021,RMP2024,HillierPRL2009}, such as CePt$_3$Si~\cite{Bauer2004,FrigeriPRL2004}, CeIrSi$_3$~\cite{MukudaPRL2008}, Li$_2$Pt$_3$B~\cite{PRLYuanHQ}, MoS$_2$~\cite{KTLaw2014} and KTaO$_{3}$~\cite{ZhaiPRL2025}. Related competition and possible admixture between $s$- and $p$-wave pairing channels have also been discussed in moiré systems~\cite{XCkPRL2018,ScheurerPRR2020,BalentsNP2020,CaoNature2021}. Yet experiments struggle to unambiguously distinguish spin-triplet pairing from an isotropic $s$-wave gap in such singlet–triplet mixed states~\cite{Smidman_2017}.
To this end, we consider a noncentrosymmetric superconductor with Rashba spin-orbit coupling (SOC) $\alpha (\boldsymbol{k} \times \boldsymbol{\sigma}) \cdot \hat{z}$, where  $\alpha$ quantifies SOC strength.
In this system, SOC forces an admixture of $s$-wave singlet and $p$-wave triplet components.
The resulting $\boldsymbol{d}$-vector locks to the SOC field in a helical pattern, $\boldsymbol{d}(\boldsymbol{k}) = k (-\sin \varphi_{\boldsymbol{k}}, \cos \varphi_{\boldsymbol{k}}, 0)$, forming an $s + \text{helical } p$ superconducting state.

As shown in Fig.~\ref{figFeNonC}(d-e), the $d_y$ component exhibits a $p$-wave symmetry on the Fermi surfaces. 
For weak spin–orbit coupling, $d_y$ is small and the pairing is $s$-wave dominated, whereas for strong spin–orbit coupling, $d_y$ is enhanced, leading to a $p$-wave–dominated state.
The nonreciprocal spin conductance $\Delta G_{s\hat{n}}$ as a function of the spatial orientation angle $\varphi$ is shown for both regimes in Fig.~\ref{figFeNonC}f. 
In both the $s$-dominated and $p$-dominated regimes, its angular dependence follows the $p$-wave pattern of $d_y$, rotated by $\pi/2$. 
Notably, singlet–triplet mixing has been proposed to occur even in the absence of SOC within moiré systems~\cite{ScheurerPRR2020,CaoNature2021}, which can also be probed by the nonreciprocal measurement, as illustrated using the toy model in Eq.~\ref{eq:BdGH} (Fig.~\ref{fignonre}f–g).

Furthermore, we complement the experimental proposal by outlining concrete schemes for probing nonreciprocal conductances, as illustrated in Fig.~\ref{figset}.
The angular dependence of $\Delta G_{c(s\hat{n})}$ can be accessed using the multi-terminal device geometry~\cite{Kang2019NonlinearAHE} shown in the upper panel of Fig.~\ref{figset}(a), where different terminals correspond to different lead--crystal orientations.
Alternatively, the same angular dependence can be obtained by rotating the crystal orientation relative to the transport direction~\cite{Bachmann2022DirectionalBallistic}. 
The measurement can also be implemented in a more flexible manner using multi-tip STM platforms~\cite{Cherepanov2012,lupke2015scanning,Lupke2017DefectResistance,Baringhaus2014BallisticGraphene,Gerasimenko2019QuantumJamming,Kolmer2019}, where the injection and detection directions are controlled by the positions of the STM tips.

Based on this setup, we now outline the measurement protocol. In particular, Fig.~\ref{figset}(b) illustrates a fixed-angle configuration.
A voltage bias $\delta V$ is applied to lead $j$, and the resulting charge (or spin) current flowing into lead $j+1$ is measured, yielding $G_{c(s\hat{n});,j+1,j}$. Exchanging the roles of the two leads gives $G_{c(s\hat{n});,j,j+1}$, and the nonreciprocal conductance is obtained from their difference, $\Delta G_{c(s\hat{n})} = G_{j+1,j} - G_{j,j+1}$.
To further reveal the phase-sensitive nature of spin-triplet pairing, we extend the protocol to spin-polarized leads [Fig.~\ref{figset}(c)]. 
When both leads are polarized along the same direction $\hat{n}$, the procedure above directly yields $\Delta G_{\uparrow_{\hat{n}},\uparrow_{\hat{n}}}$, providing a probe of $\mathcal{T}$-preserving triplet pairing. 
Replacing the pair of spin-polarized leads aligned in the same direction with a pair polarized oppositely extracts $\Delta G_{\downarrow_{\hat{n}},\uparrow_{\hat{n}}}$, which probes $\mathcal{T}$-breaking triplet pairing.


\section{Conclusion}
In summary, we establish a symmetry-resolved framework that connects superconducting pairing symmetry to nonreciprocal transport.
We show that nonreciprocal conductance provides a robust and experimentally accessible probe of unconventional pairing, capable of revealing mixed $s$–$p$ and $s$–$d$ states even when the unconventional components are masked by a dominant $s$-wave gap.
More generally, it enables the independent identification of unconventional singlet and triplet pairing components even when they coexist.

In realistic experiments, factors such as lead coupling\cite{BTK1985,CAR0,CAR1,CAR2,CAR3} and the geometric structure (e.g., interface roughness) may influence the overall conductance amplitude.
Nevertheless, they do not substantially alter the angular dependence of the nonreciprocal conductance, ensuring the robustness of unconventional pairing detection\cite{supp},
i.e., the associated interface potential fluctuations are not parametrically larger than the chemical potential~\cite{supp}.
Furthermore, the angular patterns dictated by the orbital symmetry of the superconducting order parameter provide a distinctive fingerprint of unconventional pairing, allowing it to be distinguished from other sources of nonreciprocity~\cite{supp}, such as those from edge states\cite{YasudaNN2020} or vortices\cite{RyoheiSA2017,Vortexdiode}.

This symmetry-resolved approach establishes a diagnostic criterion for hidden superconducting order by probing the pairing symmetry rather than the gap amplitude. By encoding the momentum-space structure of the pairing state into robust angular transport signatures, our framework provides a promising route to probe unconventional superconductivity across a broad class of quantum materials.

\section{Methods}
\subsection{Transmission Coefficients}
Under the low-temperature condition and the low-bias approximation
, the spin-dependent current flowing into lead $\alpha$ can be expressed as: 
\begin{equation}\label{eqLandauercurrent}
\begin{aligned}
I_{\alpha\sigma} =\frac{e^2}{h}  \sum_{\beta,\sigma'} \Big[ T^{N}_{\alpha\sigma,\beta\sigma'}(V_{\beta\sigma'}-V_{\alpha\sigma}) - T^{A}_{\alpha\sigma,\beta\sigma'}(V_{\alpha\sigma}+V_{\beta\sigma'}) \Big]
\end{aligned}
\end{equation}
with spin-dependent bias voltage $V_{\beta\sigma'}$ of terminal $\beta$ and the normal (Andreev)
transmission coefficients calculated as
$T^{N(A)}_{\alpha\sigma,\beta\sigma'}=\mathrm{Tr}\big[\mathbf{\Gamma}_{\alpha e(h)}\mathbf{G}^r\mathbf{\Gamma}_{\beta e}\mathbf{G}^a\big]$.
Here, the broadening function is defined as $\mathbf{\Gamma}_{\alpha e(h)\sigma}(\omega) = \mathrm{i} \big[ \mathbf{\Sigma}^r_{\alpha e(h)\sigma}(\omega) - \mathbf{\Sigma}^a_{\alpha e(h)\sigma}(\omega) \big]$, where $\mathbf{\Sigma}^{r,a}_{\alpha e(h)\sigma}$ are the retarded and advanced self-energies of lead $\alpha$ acting on the electron (hole) block in the BdG representation.
Through $G_{\alpha\sigma,\beta\sigma'}=\partial I_{\alpha\sigma}/\partial V_{\beta\sigma'}$ and $\Delta G_{\sigma,\sigma'} = G_{j+1\sigma,j\sigma'}-G_{j\sigma,j+1\sigma'}$, we can obtain the nonreciprocal conductance $\Delta G_{\sigma,\sigma'}$ with the transmission coefficients as shown in Eq.~\ref{eqspindepG}.
Furthermore, by the definitions of the charge current $I_{j_1c}=I_{j_1\uparrow_{\hat{n}}} +I_{j_1\downarrow_{\hat{n}}}$ and spin current $I_{j_1s\hat{n}}=\frac{\hbar}{2e}(I_{j_1\uparrow_{\hat{n}}} -I_{j_1\downarrow_{\hat{n}}})$ in terminal $\alpha$, we can obtain the nonreciprocal charge and spin conductance $\Delta G_{c(s\hat{n})}$ as shown in Eq.~\ref{eq:DGc} and Eq.~\ref{eq:DGs}.
More details of the derivation can be found in the Supplemental Material~\cite{supp}.

\subsection{Model Details}

The Hamiltonian of AFe$_2$Se$_2$ is described by $H_{\rm SC}=H_{s+\mathrm{i}d}=H_{\text{norm}} + H_{\text{pair}}$:
\begin{eqnarray} 
H_{\text{norm}} 
&=& \sum_{\mathbf{k}, \sigma} \left[ \left( \epsilon^{c}_{\mathbf{k}} - \mu \right) c^{\dagger}_{\mathbf{k}\sigma} c_{\mathbf{k}\sigma} + \left( \epsilon^{f}_{\mathbf{k}} - \mu \right) f^{\dagger}_{\mathbf{k}\sigma} f_{\mathbf{k}\sigma} \right] \nonumber \\
& & + t \left( c^{\dagger}_{\mathbf{k}\sigma} f_{\mathbf{k}\sigma} + f^{\dagger}_{\mathbf{k}\sigma} c_{\mathbf{k}\sigma} \right) , \\
H_{\text{pair}} 
&=& \sum_{\mathbf{k}, \sigma, \sigma'} (i \sigma_y)_{\sigma \sigma'}\left[ \Delta_{s} \left( c^{\dagger}_{\mathbf{k}\sigma} f^{\dagger}_{-\mathbf{k}\sigma'} + f^{\dagger}_{\mathbf{k}\sigma} c^{\dagger}_{-\mathbf{k}\sigma'} \right) \right. \nonumber \\
& & + \Delta_{d} \left( c^{\dagger}_{\mathbf{k}\sigma} c^{\dagger}_{-\mathbf{k}\sigma'} - f^{\dagger}_{\mathbf{k}\sigma} f^{\dagger}_{-\mathbf{k}\sigma'} \right) \Big] + \text{h.c.}
\end{eqnarray}
Here, \( H_{\text{norm}} \) represents the electron dispersion and $H_{\text{pair}}$ represents the pairing potential.
we consider a two-orbital model with $c$ and $f$ orbitals for $H_{\text{norm}}$, where the dispersion is given by:
$\epsilon^{c}_{\mathbf{k}} = (B+D)k_x^2+(B-D)k_y^2$
and
$\epsilon^{f}_{\mathbf{k}} = (B-D)k_x^2+(B+D)k_y^2$,
forming two orthogonally oriented elliptical Fermi surfaces.
Near the Fermi surface, we approximate the dispersion as $\epsilon^{c}_{\mathbf{k}} = \mu + v_F(\varphi_{\boldsymbol{k}})(k - k_F(\varphi_{\boldsymbol{k}}))$ and $\epsilon^{f}_{\mathbf{k}} = \mu + v_F(\varphi_{\boldsymbol{k}}+\frac{\pi}{2})(k - k_F(\varphi_{\boldsymbol{k}}+\frac{\pi}{2}))$, where $k_F(\varphi_{\boldsymbol{k}})=k_F(1 + b \cos 2\varphi_{\boldsymbol{k}})$ and $v_F(\varphi_{\boldsymbol{k}}) =v_F(1 + a \cos 2\varphi_{\boldsymbol{k}})$ are the Fermi vector and velocity, respectively.
$k_F=\sqrt{\mu/B}$, $v_F=2Bk_F$, $a=D/B$ and $b=-\frac{D}{2B}$.
The ratio of the FSs' hybridization strength to the FSs' eccentricity determines the pairing channel\cite{Khodas}.
In the absence of hybridization ($\kappa = 0$), the system hosts a pure intra-pocket $d$-wave pairing state.
The pairing channel is dictated by the ratio of the interband hybridization strength to the Fermi surface eccentricity~\cite{Khodas}. Specifically, the system evolves from a pure intra-pocket $d$-wave state at zero hybridization ($\kappa=0$) to a time-reversal symmetry-breaking $s+\mathrm{i}d$ phase, and finally to a pure interband $s$-wave state as $\kappa$ increases.

For Fig.~\ref{figFeNonC}(a-d), we set $B=\mu=1$.
Furthermore, the pairing strengths $(\Delta_s, \Delta_d)$ and the corresponding coefficient $\kappa=t/(v_Fk_F|b|)$ for four representative cases are: (i) pure $s$-wave: $(0.15, 0)\mu$ with $\kappa=5$; (ii) pure $d$-wave: $(0, 0.15)\mu$ with $\kappa=0.3$; (iii) $s$-dominant $s+\mathrm{i}d$: $(0.1, 0.05\mathrm{i})\mu$ with $\kappa=2$; and (iv) $d$-dominant $s+\mathrm{i}d$: $(0.05, 0.1\mathrm{i})\mu$ with $\kappa=0.5$.
$D$ here is set to be positive.

For the noncentrosymmetric superconductor, we apply $H_{\text{SC}}=H_{\text{nonc}}$\cite{Gorkov2001,Sigrist2004}
\begin{align} 
H_{\text{nonc}}={}
&\sum_{\mathbf{k},\sigma}\bigl[\epsilon(\boldsymbol{k}) - \mu\bigr]c^{\dagger}_{\mathbf{k}\sigma}c_{\mathbf{k}\sigma}+\alpha\sum_{\sigma,\sigma',\boldsymbol{k}}\boldsymbol{g}(\boldsymbol{k})\cdot\boldsymbol{\sigma}_{\sigma,\sigma'}c^{\dagger}_{\mathbf{k}\sigma}c_{\mathbf{k}\sigma'}\nonumber\\ 
&+\bigl[\Delta_s\sum_{\sigma,\sigma',\boldsymbol{k}}(\mathrm{i}\sigma_y)_{\sigma,\sigma'}c^{\dagger}_{\mathbf{k}\sigma}c^{\dagger}_{-\mathbf{k}\sigma'}.\nonumber\\ &+\Delta_p\sum_{\sigma,\sigma',\boldsymbol{k}}\boldsymbol{\tilde{d}}(\boldsymbol{k})\cdot(\mathrm{i}\boldsymbol{\sigma}\sigma_y)_{\sigma,\sigma'}c^{\dagger}_{\mathbf{k}\sigma}c^{\dagger}_{-\mathbf{k}\sigma'}+h.c.\bigr]. 
\label{eqH_s_id} 
\end{align}
Here, $\boldsymbol{g}(\boldsymbol{k})$ characterizes the spin-orbit coupling (SOC) with a strength parameter $\alpha$, which locks the spin-triplet $\boldsymbol{\tilde{d}}$-vector parallel to $\boldsymbol{g}(\boldsymbol{k})$~\cite{Sigrist2004}.
Specifically, for Rashba-type SOC, $\boldsymbol{\tilde{d}}(\boldsymbol{k}) \propto \boldsymbol{g}(\boldsymbol{k}) = k \left( -\sin(\varphi_{\boldsymbol{k}}), \cos(\varphi_{\boldsymbol{k}}), 0 \right)$, thereby establishing a helical \( p \)-wave.
A quadratic dispersion relation $\epsilon({\boldsymbol{k}})=Bk^2$ with $B=1$, $\mu=1$ and $\Delta_s=0.2$ is adopted for Fig.~\ref{figFeNonC}(d-f).
Furthermore, $\Delta_p = 0.03$ for the $s$-wave dominant case with $\alpha = 0.1$, while $\Delta_p = 0.3$ for the $p$-wave dominant case with $\alpha = 1$.

To compute the nonreciprocal conductance numerically, we employ a finite-difference tight-binding lattice model.
More details can be found in the Supplemental Material~\cite{supp}.

\subsection{Effective pairing on the Fermi Surface}
The effective pairing on the Fermi surface shown by Fig.~\ref{figFeNonC}(a,b,d,e) is obtained by projecting the pairing potential onto the Fermi surface states as
\begin{align}\label{eqProj}
[\hat{\Delta}^{n}_{\text{eff}}(\boldsymbol{k})]_{\sigma,\sigma'} = \psi_{\boldsymbol{k}\sigma,n}\psi_{-\boldsymbol{k}\sigma',n}\sum_{\sigma_1,\sigma_2} \psi^{*}_{\boldsymbol{k}\sigma_1,n}[\hat{\Delta}(\boldsymbol{k})]_{\sigma_1,\sigma_2} \psi^{*}_{-\boldsymbol{k}\sigma_2,n}
\end{align}
with $\boldsymbol{k}\in$  the Fermi surface of the $n$-th band.
So $ \psi_{\boldsymbol{k}\sigma,n} $ is the normal-state eigenstate of the $n$-th band at the Fermi surface.
$\hat{\Delta}$ is the pairing-potential matrix in the orbital basis.
By decomposing $\hat{\Delta}^{n}_{\text{eff}}$ into spin-singlet and triplet channels, we can obtain the effective order parameter $\psi$ for spin-singlet pairing and the $\boldsymbol{d}$-vector for spin-triplet pairing on the Fermi surface as
\begin{equation}
\hat{\Delta}^{n}_{\text{eff}} = \mathrm{i} \Big[ \psi(\boldsymbol{k}) + \boldsymbol{d}(\boldsymbol{k})\cdot \boldsymbol{\sigma}\Big] \sigma_y.
\label{eq:effgapmatrix}
\end{equation}

\section*{Data Availability}
The data that support the findings of this study are available from the corresponding author upon reasonable request.
\\
\section*{Acknowledgment}
We thank Qing-Feng Sun, Humian Zhou, Yue Mao, Xilin Feng for illuminating discussions.
This work was financially supported by National Key R and D Program of China (Grants No. 2022YFA1403700, 2017YFA0303301), NSFC (Grants Nos. 11534001, 11822407, 11921005, 12074108), and also supported
by the Fundamental Research Funds for the Central Universities, the Strategic Priority Research Program of Chinese Academy of Sciences (DB28000000), and Beijing Municipal Science \& Technology
Commission (Grant No. Z191100007219013).
C.-Z. C. was also supported by the Natural Science Foundation of Jiangsu Province Grant (No. BK20230066),  Jiangsu Shuang Chuang Project (JSSCTD202209) and the Priority Academic Program Development (PAPD) of Jiangsu Higher Education Institution.
Ming Gong is also funded by China Postdoctoral Science Foundation (Grant No. BX20240004).
\\
\section*{Author Contributions}
X. C. Xie. and C. -Z. Chen conceived the idea and initiated the project. W. -B. Dai. performed the theoretical calculations. All the authors discussed the results and co-wrote the paper.
\\
\section*{Competing Interests}
The authors declare no competing interests.

\bibliography{shiftref}

\bibliographystyle{apsrev4-2}      

\end{document}


\title{Supplementary Materials for ``Revealing Hidden Unconventional Pairing through Nonreciprocal Transport"}
\author{Wen-Bo Dai}
\affiliation{International Center for Quantum Materials, School of Physics, Peking University, Beijing 100871, China}
\affiliation{Beijing Academy of Quantum Information Sciences, Beijing 100193, China}
\affiliation{Department of Physics, Hong Kong University of Science and Technology, Clear Water Bay, Hong Kong, China}
\author{Ming Gong}
\affiliation{International Center for Quantum Materials, School of Physics, Peking University, Beijing 100871, China}
\affiliation{Department of Physics, The University of Tokyo, 7-3-1 Hongo, Tokyo 113-0033, Japan}
\author{Xianxin Wu}
\affiliation{CAS Key Laboratory of Theoretical Physics, Institute of Theoretical Physics,
Chinese Academy of Sciences, Beijing 100190, China}
\author{Chui-Zhen Chen}
\email{czchen@suda.edu.cn}
\affiliation{School of Physical Science and Technology, Soochow University, Suzhou 215006, China}
\affiliation{Institute for Advanced Study, Soochow University, Suzhou 215006, China}
\author{X. C. Xie}
\email{xcxie@pku.edu.cn}
\affiliation{International Center for Quantum Materials, School of Physics, Peking University, Beijing 100871, China}
\affiliation{Interdisciplinary Center for Theoretical Physics and Information Sciences, Fudan University, Shanghai 200433, China}
\affiliation{Hefei National Laboratory, Hefei 230088, China}
\date{\today }

\maketitle
\tableofcontents

\section{ General Landauer's Transport Formalism in Superconducting Systems}
In the seminal work of Y. Meir and N. S. Wingreen\cite{Wingreen1992,Wingreen1993},
the Landauer formula has been extended to an interacting electron system. 
Following this seminal work,  the Landauer formula was subsequently applied to superconducting systems \cite{Sun2000}, where it was demonstrated to account for the contribution from Andreev (normal) reflection, as it includes scattering between electron and hole (electron) channels.
Below, we explicitly extend the Landauer formula to superconducting systems.
Following Refs.~\cite{Wingreen1992,Wingreen1993} and \cite{Sun2000,Sun_2009},
we consider a superconducting center region coupled with normal leads and a superconducting lead. 
The spin-polarized current $I_{\alpha\sigma}$ with spin $\sigma$ flows into the $\alpha$-th normal lead is given by \cite{Wingreen1992}:
\begin{align}
I_{\alpha\sigma}
&= -e\left\langle \frac{d \hat{N}_{\alpha\sigma}}{dt} \right\rangle \notag \\
&= \frac{e}{h} \int d\omega\,
\mathrm{Tr}\Big\{
\boldsymbol{\Gamma}_{\alpha e\sigma}(\omega)
\Big[
f_{\alpha\sigma+}(\omega)
\big(
\mathbf{G}^r(\omega) - \mathbf{G}^a(\omega)
\big)
+ \mathbf{G}^<(\omega)
\Big]
\Big\}.
\end{align}

Here $\hat{N}_{\alpha\sigma}$ is the number operator for electrons with spin $\sigma$ in the $\alpha$-th lead.
The matrices $\mathbf{G}^{r,~a,~<}(\omega)$ represent the Fourier transformation of the retarded, greater and lesser Green function matrix in the BdG representation, respectively.
The distribution functions are defined as $f_{\alpha\sigma\pm}(\omega)
= f(\omega \mp \tilde{\mu}_{\alpha\sigma})
= f(\omega \pm e V_{\alpha\sigma})$ for electron or hole, where $\tilde{\mu}_{\alpha\sigma}$ is the
electrochemical potential and $V_{\alpha\sigma}$ is the applied voltage for
spin $\sigma$ in lead $\alpha$. The Fermi--Dirac distribution is given by
$f(\omega) = 1/(e^{\beta \omega} + 1)$, with $\beta = 1/(k_B T)$.
Meanwhile, for the superconducting lead, we have $f_{S}(\omega)=f(\omega-\mu_S)=f(\omega+eV_{S})$.
The broadening function is defined by 
$\mathbf{\Gamma}_{\alpha e(h)\sigma}(\omega)=\mathrm{i}\big[\mathbf{\Sigma}^r_{\alpha e(h)\sigma}(\omega)-\mathbf{\Sigma}^a_{\alpha e(h)\sigma}(\omega)\big]$, 
where the self-energies take the form 
$\mathbf{\Sigma}^{r,a}_{\alpha e(h)\sigma}=H_T\,\mathbf{g}^{r,a}_{\alpha e(h)\sigma}\,H_T$. 
Here $\mathbf{g}^{r,a}_{\alpha e(h)\sigma}$ denote the electron (hole) Green’s functions of the $\sigma$ channel in the $\alpha$-th lead, 
and $H_T$ represents the coupling Hamiltonian between the central region and the leads.
Similarly, for the superconducting lead one has 
$\mathbf{\Gamma}_{S}(\omega)=\mathrm{i}\big[\mathbf{\Sigma}^r_{S}(\omega)-\mathbf{\Sigma}^a_{S}(\omega)\big]$ 
with the self energy $\mathbf{\Sigma}^{r,a}_{S}=H_T\,\mathbf{g}^{r,a}_{S}\,H_T$.
The lesser self-energy associated with the coupling to the $\sigma$ channel of the $\alpha$-th normal lead is given by
\begin{eqnarray}\label{eqSigma2}
	\mathbf{\Sigma}^{<}_{\alpha\sigma}(\omega)&=&\mathbf{\Sigma}^{<}_{\alpha e\sigma}(\omega)\oplus\mathbf{\Sigma}^{<}_{\alpha h\sigma}(\omega)\nonumber\\
	&=&\mathrm{i}(f_{\alpha\sigma +}(\omega) \mathbf{\Gamma}_{\alpha e\sigma}(\omega) \oplus  f_{\alpha\sigma -}(\omega)\mathbf{\Gamma}_{\alpha h\sigma}(\omega))
\end{eqnarray}
where $\mathbf{\Sigma}^{<}_{\alpha e(h)\sigma}(\omega)$ denote the electron (hole) components, respectively. 
For the superconducting lead, the lesser self-energy reads
\begin{eqnarray}\label{eqSigma3} 
\mathbf{\Sigma}^{<}_S(\omega)=\mathrm{i}f_{S}(\omega)\mathbf{\Gamma}_{S}(\omega).
\end{eqnarray}
Then by using the Keldysh equation $\mathbf{G}^{<}=\mathbf{G}^{r}\mathbf{\Sigma}^{<}\mathbf{G}^{a}$ and $\mathbf{G}^r-\mathbf{G}^a=\mathbf{G}^r(\mathbf{\Sigma}^r-\mathbf{\Sigma}^a)\mathbf{G}^a$,
we obtain the spin-polarized current expression:
\begin{eqnarray}\label{eqLandauer2}
	I_{\alpha\sigma}&=&\frac{e}{h}\int d\omega\{\sum_{\beta,\sigma'}[(f_{\alpha\sigma+}(\omega)-f_{\beta\sigma'+}(\omega))T^{N}_{\alpha\sigma,\beta\sigma'}(\omega)+(f_{\alpha\sigma+}(\omega)-f_{\beta\sigma'-}(\omega))T^{A}_{\alpha\sigma,\beta\sigma'}(-\omega)]\nonumber\\
	&~&+(f_{\alpha\sigma+}(\omega)-f_{S}(\omega))T_{\alpha\sigma e,S}(\omega)\}
\end{eqnarray}
Here, $T^{N(A)}_{\alpha\sigma,\beta\sigma'}=\mathrm{Tr}\big[\mathbf{\Gamma}_{\alpha e(h)}\mathbf{G}^r\mathbf{\Gamma}_{\beta e}\mathbf{G}^a\big]$ represents the normal/Andreev transmission probability.
Meanwhile, $T_{\alpha e,S}=\mathrm{Tr}\big[\mathbf{\Gamma}_{\alpha e}\mathbf{G}^r\mathbf{\Gamma}_{S}\mathbf{G}^a\big]$ denotes the transmission from the superconducting lead to the $\alpha$-th normal lead.
Therefore, the generalized Landauer formula derived here naturally incorporates contributions from both Andreev and normal reflections through the terms $T^{A}$ and $T^{N}$ in superconducting systems.

Under the low-temperature condition $f_{\beta\sigma'\pm}(\omega)=\theta(\mp eV_{\beta\sigma'}-\omega)$ and the low-bias approximation  $V_{\beta \sigma'}\sim 0$
, the above formalism in Eq.\ref{eqLandauer2} simplifies to 
\begin{eqnarray}\label{eqLandauer3}
	I_{\alpha\sigma}&=&\frac{e^2}{h}\{\sum_{\beta}[T^{N}_{\alpha\sigma,\beta\sigma'}(V_{\beta\sigma'}-V_{\alpha\sigma})-T^{A}_{\alpha\sigma,\beta\sigma'}(V_{\alpha\sigma}+V_{\beta\sigma'})]+T_{\alpha\sigma e,S}(V_{S}-V_{\alpha\sigma})\}
\end{eqnarray}
When the superconducting systems is coupled solely to normal leads, one can get Eq.~(11) of the manuscript. 
The spin-dependent conductance $G_{\alpha\sigma,\beta\sigma'}=\frac{\partial I_{\alpha\sigma}}{\partial V_{\beta\sigma'}}$ in response to the spin-polarized bias $V_{\beta\sigma'}$ is then given by
\begin{eqnarray}\label{eqG1}
	G_{\alpha\sigma,\beta\sigma'}=\frac{e^2}{h}(T^{N}_{\alpha\sigma,\beta \sigma'}-T^{A}_{\alpha\sigma,\beta \sigma'})
\end{eqnarray}
for $\alpha\neq\beta$.

Furthermore, the charge (spin) conductance 
$G_{c(s\hat{n})}=I_{c(s\hat{n})}/V_{\beta}$
in response to a charge bias $V_{\beta\sigma'}=V_{\beta}$ applied to the $\beta$th terminal
is expressed as
\begin{eqnarray}\label{eqG2}
	\left\{
	\begin{array}{lll}
        G_{\alpha c,\beta}&=&\sum_{\sigma,\sigma'} G_{\sigma,\beta\sigma'} \\
        G_{\alpha s\hat{n},\beta}&=&\frac{\hbar}{2e} \sum_{\sigma'}(G_{\uparrow_{\hat{n}},\beta\sigma'} - G_{\downarrow_{\hat{n}},\beta\sigma'} )
    \end{array}
	\right.
\end{eqnarray}
Here, Eq.~(\ref{eqG2}) follows directly from the definitions of the charge and spin currents,
$I_{\alpha c}=I_{\alpha\uparrow_{\hat{n}}}+I_{\alpha\downarrow_{\hat{n}}}$
and
$I_{\alpha s\hat{n}}=\frac{\hbar}{2e}(I_{\alpha\uparrow_{\hat{n}}}-I_{\alpha\downarrow_{\hat{n}}})$, respectively.
Using Eq.~(\ref{eqG2}), one can further establish the relation between the spin-dependent nonreciprocal conductances and the nonreciprocal charge (spin) conductances, as summarized in Eqs.~(5–6) of the manuscript.

\section{Derivation of Nonreciprocal Conductance}
Using the Dyson equation, $\mathrm{G}^{r(a)}=\mathrm{g}^{r(a)}+\lambda\,\mathrm{g}^{r(a)}V_{un}\mathrm{g}^{r(a)}+\dots$,
the conductance $G_{\alpha\sigma,\beta\sigma'}~(\alpha\neq\beta)$can be expanded up to the first order in $\lambda$ as
\begin{eqnarray}\label{eqGdyson}
G_{\alpha\sigma,\beta\sigma'} &=&
2\,\mathrm{Re}\!\Big(\mathrm{Tr}\!\left[
  (\mathbf{\Gamma}_{\alpha e\sigma} - \mathbf{\Gamma}_{\alpha h\sigma})
  \mathbf{g}^r\mathbf{\Gamma}_{\beta e\sigma'} \mathbf{g}^{a}
\right]\!\Big)
+ 2\lambda\,\mathrm{Re}\!\Big(\mathrm{Tr}\!\left[
  (\mathbf{\Gamma}_{\alpha e\sigma} - \mathbf{\Gamma}_{\alpha h\sigma})
  \mathbf{g}^r V_{un}(\varphi_{\mathbf{k}}) \mathbf{g}^r \mathbf{\Gamma}_{\beta e\sigma'} \mathbf{g}^{a}
\right]\!\Big),
\end{eqnarray}
by expanding the transmission probability as
\begin{eqnarray}\label{eqTdyson}
T^{N(A)}_{\alpha\sigma,\beta\sigma'} & = &
2\,\mathrm{Re}\!\Big(\mathrm{Tr}\!\left[
  \mathbf{\Gamma}_{\alpha e(h)\sigma}\mathbf{g}^r\mathbf{\Gamma}_{\beta e\sigma'}\mathbf{g}^a
\right]\!\Big)
+ 2\lambda\,\mathrm{Re}\!\Big(\mathrm{Tr}\!\left[
  \mathbf{\Gamma}_{\alpha e(h)\sigma}
  \mathbf{g}^r V_{un}(\varphi_{\mathbf{k}}) \mathbf{g}^r \mathbf{\Gamma}_{\beta e\sigma'}\mathbf{g}^a
\right]\!\Big).
\end{eqnarray}
in Eq.\ref{eqG1}.
Here  $\mathbf{G}^{r(a)}$ and $\mathbf{g}^{r(a)}$ denote the retarded (advanced) Green’s functions of the total and unperturbed systems, respectively.

\begin{figure}[bht]
    \centering
    \includegraphics[width=6.6in]{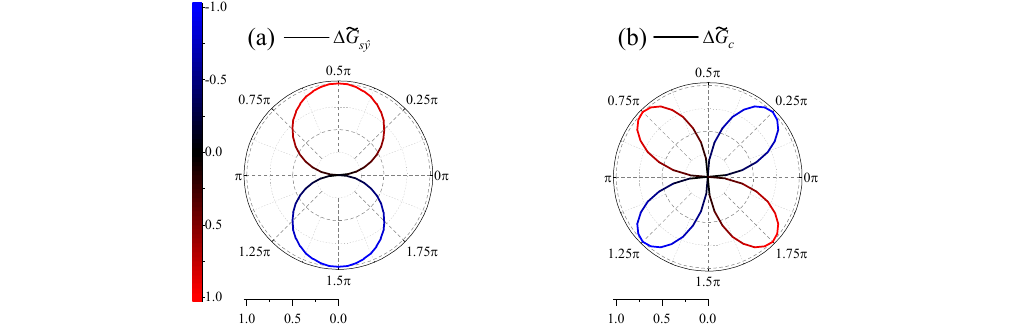}
    \caption{(Color online).
	Polar plots of  the normalized nonreciprocal conductances $\Delta\Tilde{G}(\varphi)\equiv\Delta G(\varphi)/\mathrm{max}(|\Delta G(\varphi)|)$ versus angle $\varphi$ for mixed $s-p-d$ pairing.
    with $\boldsymbol{d}=-\hat{x}\mathrm{sin}(\varphi_{\boldsymbol{k}})+\hat{y}\mathrm{cos}(\varphi_{\boldsymbol{k}})$ and 
	$\Psi_{un}=\mathrm{icos}(2\varphi_{\boldsymbol{k}})$
	(a) $\Delta\Tilde{G}_{s\hat{y}}$ 
    (b) $\Delta\Tilde{G}_{c}$ versus $\varphi$ . The parameters are $B = 1$, $\mu = 1$, $\Delta_s = 0.2$ and $\lambda = 0.02$.
	\label{FigSspd} }
\end{figure}

To analyze the nonreciprocal conductance, we consider two adjacent leads, $j$ and $j+1$, located around the normal angle $\varphi$ [see Fig.~1(a) of the manuscript].
By evaluating $G_{j+1\sigma,j\sigma'}$ and $G_{j\sigma,j+1\sigma'}$, the spin-dependent nonreciprocal conductance can be obtained from their difference,
\begin{align}\label{eqDGdyson} 
\Delta G_{\sigma,\sigma'}(\varphi)
&= G_{j+1\sigma,j\sigma'} - G_{j\sigma,j+1\sigma'} \nonumber\\
&= 2\lambda\,\mathrm{Re}\!\Big\{
\mathrm{Tr}\!\Big[
  (\mathbf{\Gamma}_{j+1e\sigma} - \mathbf{\Gamma}_{j+1h\sigma})
  \mathbf{g}^r V_{un}(\varphi_{\mathbf{k}}) \mathbf{g}^r \mathbf{\Gamma}_{je\sigma'}
  - (\mathbf{\Gamma}_{je\sigma} - \mathbf{\Gamma}_{jh\sigma})
  \mathbf{g}^r V_{un}(\varphi_{\mathbf{k}}) \mathbf{g}^r \mathbf{\Gamma}_{j+1e\sigma'}
\Big]
\mathbf{g}^a
\Big\}\nonumber\\
&=2\lambda\,\mathrm{Re}\!\Big\{
\mathrm{Tr}\!\Big[
  (\mathbf{\Gamma}_{j+1e\sigma} - \mathbf{\Gamma}_{j+1h\sigma})
  \mathbf{g}^r V_{un}(\varphi_{\mathbf{k'}}+\varphi) \mathbf{g}^r \mathbf{\Gamma}_{je\sigma'}
  - (\mathbf{\Gamma}_{je\sigma} - \mathbf{\Gamma}_{jh\sigma})
  \mathbf{g}^r V_{un}(\varphi_{\mathbf{k}}+\varphi) \mathbf{g}^r \mathbf{\Gamma}_{j+1e\sigma'}
\Big]
\mathbf{g}^a
\Big\}\nonumber\\
&=2\lambda\,\mathrm{Re}\!\Big\{
\mathrm{Tr}\!\Big[
  (\mathbf{\Gamma}_{j+1e\sigma} - \mathbf{\Gamma}_{j+1h\sigma})
  \mathbf{g}^r \big( V^{l}_{un}(\varphi+\tfrac{\pi}{2l}) \sin(l\varphi_{\mathbf{k'}})
   + V^{l}_{un}(\varphi)\cos(l\varphi_{\mathbf{k'}}) \big) \mathbf{g}^r \mathbf{\Gamma}_{je\sigma'}\nonumber\\
&\qquad\qquad\qquad-(\mathbf{\Gamma}_{je\sigma} - \mathbf{\Gamma}_{jh\sigma})
  \mathbf{g}^r \big( V^{l}_{un}(\varphi+\tfrac{\pi}{2l}) \sin(l\varphi_{\mathbf{k'}})
   + V^{l}_{un}(\varphi)\cos(l\varphi_{\mathbf{k'}}) \big) \mathbf{g}^r \mathbf{\Gamma}_{j+1e\sigma'}
\Big]
\mathbf{g}^a
\Big\}\nonumber\\
&=2\lambda\,\mathrm{Re}\!\Big\{
\mathrm{Tr}\!\Big[
  (\mathbf{\Gamma}_{2e\sigma} - \mathbf{\Gamma}_{2h\sigma})
  \mathbf{g}^r V^{l}_{un}(\varphi+\tfrac{\pi}{2l}) \sin(l\varphi_{\mathbf{k}}) \mathbf{g}^r \mathbf{\Gamma}_{1e\sigma'}\nonumber\\
&\qquad\qquad\qquad-(\mathbf{\Gamma}_{1e\sigma} - \mathbf{\Gamma}_{1h\sigma})
  \mathbf{g}^r V^{l}_{un}(\varphi+\tfrac{\pi}{2l}) \sin(l\varphi_{\mathbf{k}}) \mathbf{g}^r \mathbf{\Gamma}_{2e\sigma'}
\Big]
\mathbf{g}^a
\Big\}.
\end{align}
This second line corresponds to Eq.~(5) of the manuscript, where the zeroth-order term in $\lambda$ vanishes due to the reciprocal nature of the unperturbed system.  
In the third line, $\mathbf{k'}$ denotes the momentum in the rotated $x'\!-\!y'$ frame associated with the normal angle $\varphi$.  
The quantity $V^{l}_{\mathrm{un}}(\theta)$ denotes the $l$-fold symmetric (i.e., $l$-th harmonic) component of the angular-dependent potential $V_{\mathrm{un}}(\theta)$, defined through the standard Fourier–harmonic projection:
\[
V^{l}_{\mathrm{un}}(\theta)
=\frac{1}{\pi}
\int_{-\pi}^{\pi} d\theta'
\,V_{\mathrm{un}}(\theta')
\cos\!\big(l(\theta'-\theta)\big)
= A_l\cos(l\theta)+B_l\sin(l\theta),
\]
with
\[
A_l=\frac{1}{\pi}\!\int_{-\pi}^{\pi}\!d\theta'\,V_{\mathrm{un}}(\theta')\cos(l\theta'),
\qquad
B_l=\frac{1}{\pi}\!\int_{-\pi}^{\pi}\!d\theta'\,V_{\mathrm{un}}(\theta')\sin(l\theta').
\]
In the sixth-seventh line, we assume that the leads are rotationally symmetric. Even if this symmetry is moderately broken, it would only introduce minor errors in the angular resolution.
According to the unconventional pairing potential given in Eq.~(6) of the manuscript, the spin-dependent nonreciprocal conductance $\Delta G_{\sigma,\sigma'}$ is constrained by
\begin{eqnarray}\label{eqkappa2}
    \Delta G_{\sigma,\sigma'}(\varphi)&=& \lambda\sum_{l}\mathrm{Im}(\chi_{l;\sigma,\sigma'}d^{l}_n(\varphi+\frac{\pi}{2l})+\xi_{l;\sigma,\sigma'}\psi^{l}_{un}(\varphi+\frac{\pi}{2l}))
\end{eqnarray}
for a chosen spin-axis $\hat{n}$ with $\sigma,\sigma'\in\{\uparrow_{\hat{n}},\downarrow_{\hat{n}}\}$.
Here we assume that the \( s \)-wave pairing parameter \(\Delta_s\) is purely real, in accordance with the conventions used in the manuscript.
Besides, $d_{n}^l$ ($\psi_{un}^{l}$) represents the $l$-fold symmetric part of $d_{n}$ ($\psi_{un}$), in the same manner as $V^{l}_{un}$.
The expressions for $\boldsymbol{\chi}_{l;\sigma,\sigma'}$ and $\xi_{l;\sigma,\sigma'}$ are given by:
\begin{eqnarray}\label{eqchi}
    \chi_{l;\sigma,\sigma'}&=&-\mathcal{D}_{\sigma,\sigma'}(\tau_+ \sigma_n\sigma_y\mathrm{sin}(l\varphi_{\mathbf{k}}))+\mathrm{i}\mathcal{D}_{\sigma,\sigma'}(i\tau_+ \sigma_n\sigma_y\mathrm{sin}(l\varphi_{\mathbf{k}}))
\end{eqnarray}
and
\begin{eqnarray}\label{eqxi}
    \xi_{l;\sigma,\sigma'}&=&-\mathcal{D}_{\sigma,\sigma'}(\tau_+\sigma_y \mathrm{sin}(l\varphi_{\mathbf{k}}))
\end{eqnarray}
Here we define $\mathcal{D}_{\sigma,\sigma'}$ as
\begin{equation}
\mathcal{D}_{\sigma,\sigma'}(\mathrm{A}) = 2\mathrm{Re}\left(\mathrm{Tr} \Big\{ \left[ (\mathbf{\Gamma}_{2e\sigma} - \mathbf{\Gamma}_{2h\sigma}) \mathbf{g}^r (\mathrm{A} + \mathrm{A}^{\dagger}) \mathbf{g}^r \mathbf{\Gamma}_{1e\sigma'} \right. - \left. (\mathbf{\Gamma}_{1e\sigma} - \mathbf{\Gamma}_{1h\sigma}) \mathbf{g}^r (\mathrm{A} + \mathrm{A}^{\dagger}) \mathbf{g}^r \mathbf{\Gamma}_{2e\sigma'} \right] \mathbf{g}^a \Big\}\right)
\label{eq:Dyson}
\end{equation}
Here, the pair of leads $1$, $2$ are located around the zero normal angle [see Fig.1(a) in the manuscript].
We assume that lead $j(j+1)$ located around normal angle $\varphi$ can be generated by rotating lead $1(2)$ through $\mathrm{SO}(2)$ transformation as $\Gamma_{j(j+1)}=\mathbf{U_R}(\varphi)\Gamma_{1(2)}\mathbf{U_R^{\dagger}}(\varphi)$.
When the orbital angular momentum \( l =2n\) is even, i.e., $\psi^{l=2n}_{un} \neq 0$ and all other components vanish,
we recover Eq.~(7) of the main text.
Conversely, when \( l=2n+1 \) is odd, i.e., $\boldsymbol{d}^{l=2n+1} \neq 0$ and all other components vanish,
we obtain Eqs. (8-9) of the main text.
Furthermore, we verify the applicability of the nonreciprocal conductance method in probing a more complex unconventional pairing.
To achieve this,  we consider a scenario involving mixed \( s \)-\( p \)-\( d \) pairing ($s+$ helical $p+$ $\mathrm{i}d$).
According to Eq. (7) and Eq. (9) in the manuscript, the nonreciprocal charge and spin conductances can independently probe \( p \)-wave and \( d \)-wave pairing.
As a result, the nonreciprocal spin and charge conductances reveal the \( p \)-wave and \( d \)-wave patterns in Fig.\ref{FigSspd}(a-b), respectively.
In addition, in the most general case, when unconventional superconductivity involves an arbitrary mixture of different orbital angular momenta, the Fourier transform of the nonreciprocal conductances with respect to the normal angle $\varphi$ allows one to extract the components corresponding to orbital angular momentum $l$ in the unconventional pairing, based on Eq.~\ref{eq:combined} and its $l$-fold rotational symmetry.

\section{Symmetry Constraints and Reciprocity}
In this section, we demonstrate the reciprocity of scattering channels under the corresponding symmetries.

For a Hamiltonian with time-reversal symmetry \(\mathcal{T}\), which satisfies
\begin{equation} \label{eqHT}
	[H, \mathcal{T}] = 0,
\end{equation}
the scattering matrix \(S\) satisfies the relation
\begin{equation} \label{eqST}
	S_{\psi_{n},\psi_m} = S_{\mathcal{T}\psi_m, \mathcal{T}\psi_n}.
\end{equation}
Here, the left (right) index corresponds to the outgoing (incoming) wave $\Psi^{\mp}$, 
and the scattering matrix $S$ relates the incoming states 
$\Psi^{+} = [\psi^{+}_1, \dots, \psi^{+}_n, \dots, \psi^{+}_N]$ 
to the outgoing states 
$\Psi^{-} = [\psi^{-}_1, \dots, \psi^{-}_n, \dots, \psi^{-}_N]$.
Besides, $\mathcal{T}\Psi_n$ denotes the time-reversed states $\mathcal{T}\psi^{\mp}_n$, corresponding to the incoming (outgoing) channels when appearing as the right (left) index of the scattering matrix $S$.
Eq.~\ref{eqST} implies that both the scattering direction and the orbital motion are reversed under the time-reversal transformation.
According to the scattering matrix $S$, the eigenstate corresponding to energy \( E \) in the leads is given by \( \Phi = \Psi^{+} A + \Psi^{-} \mathbf{S} A \), where \([\mathbf{S}]_{nm} = S_{\psi_n, \psi_m}\) and \( A = [a_1, \dots, a_n, \dots, a_N]^T \) representing incoming amplitudes.
Under time-reversal symmetry, as described by Eq. \ref{eqHT}, the time-reversal state \( \mathcal{T} \Phi \) remains an eigenstate with energy \( E \), and is expressed as \( \mathcal{T} \Phi = (\mathcal{T} \Psi^+) A^* + (\mathcal{T} \Psi^{-}) \mathbf{S}^* A^* \).
By defining \( A' = \mathbf{S}^* A \), we can rewrite \( \mathcal{T} \Phi \) as \( \mathcal{T} \Phi = (\mathcal{T} \Psi^+) \mathbf{S}^T A' + (\mathcal{T} \Psi^{-}) A' \), considering that \( \mathbf{S}^{\dagger} = \mathbf{S}^{-1} \) due to probability conservation..
Thus, we arrive at the relation \( S_{\mathcal{T} \psi_m, \mathcal{T} \psi_n} = [\mathbf{S}^T]_{mn} = S_{\psi_{n}, \psi_{m}} \), as stated in Eq. \ref{eqST}.

Considering both electron/hole and spin channels, we derive the following relations:
\begin{equation} \label{eq:combined}
	\left\{
	\begin{array}{l}
		S_{\alpha \sigma e, \beta \sigma' e} = -S_{\beta \overline{\sigma'} e, \alpha \overline{\sigma} e}, \\
		S_{\alpha \sigma h, \beta \sigma' e} = -S{^*}_{\beta \overline{\sigma'} h, \alpha \overline{\sigma} e}.
	\end{array}
	\right.
\end{equation}
Here, \(e~(h)\) denotes electron (hole), and \(\sigma, \sigma'\) represent spin indices.
Besides, \(\overline{\sigma}\) denotes the state with reversed spin of $\sigma$ state.
In the second line of Eq.\ref{eq:combined}, we take into account the particle-hole symmetry of the Bogoliubov-de Gennes (BdG) Hamiltonian.
For spin-flip scattering processes with \(\sigma = \overline{\sigma'}\), we obtain the relation:
\begin{equation} \label{eqTre1}
	T^{N(A)}_{\alpha \overline{\sigma}, \beta \sigma} = T^{N(A)}_{\beta \overline{\sigma}, \alpha \sigma}.
\end{equation}
Here, \(T^{N(A)}_{\alpha \overline{\sigma}, \beta \sigma} = |S_{\alpha \overline{\sigma} e(h), \beta \sigma e}|^2\) represents the transmission probability for normal (Andreev) spin-flip scattering from terminal \(\beta\) to terminal \(\alpha\).
Eq.\ref{eqTre1} demonstrates that both types of spin-flip scattering processes obey reciprocity under time-reversal symmetry \(\mathcal{T}\).

Similarly, for the system preserving the spin-time combined symmetry $s_{\hat{n}_\perp}\mathcal{T}$, we have
\begin{equation} \label{eqTre2}
	T^{N(A)}_{\alpha \sigma, \beta \sigma} = T^{N(A)}_{\beta \sigma, \alpha \sigma}.
\end{equation}
Eq.\ref{eqTre2} demonstrates that both types of equal-spin scattering processes obey reciprocity under spin-time symmetry \(s_{\hat{n}_\perp}\mathcal{T}\).

Thus, the pure $s$-wave pairing does not lift either the time-reversal constraint $\mathcal{T}$ governing spin-flip scattering or the combined spin-time-reversal constraint $s_{\hat{n}_{\perp}}\mathcal{T}$ governing equal-spin scattering.
The corresponding normal and Andreev scattering probabilities therefore remain reciprocal, and the charge and spin nonreciprocal responses associated with these channels are filtered out.
Importantly, although geometrical asymmetry can break mirror symmetry, it does not lift either of these scattering-channel reciprocity constraints.
Consequently, geometrical asymmetry cannot generate nonreciprocity from the pure $s$-wave pairing contribution.
\section{Numerical Verifications of the Spin-Resolved Microscopic Mechanism}
\label{sec:sm_verification}
\begin{figure*}[bht]
    \centering
    \includegraphics[width=6.8in]{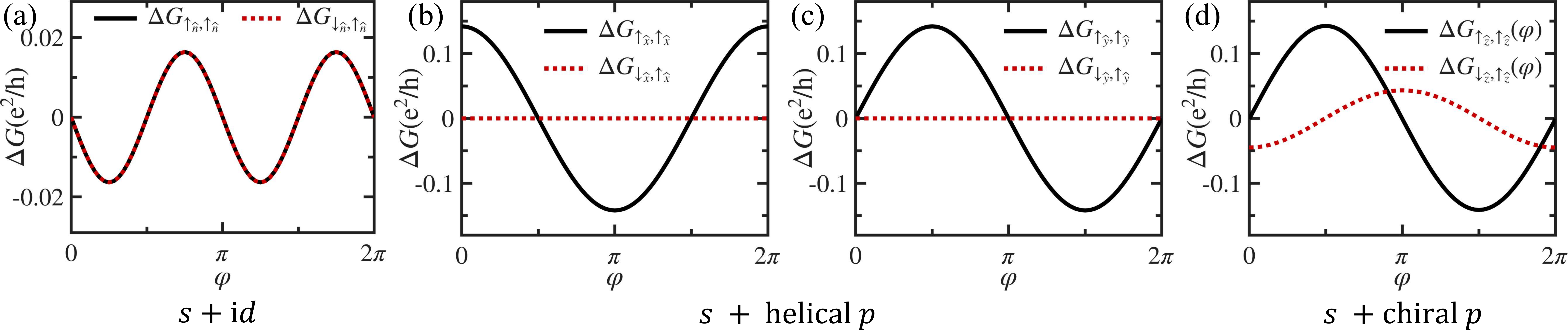}
    \caption{(Color online).
	(a-d) Polar plots of  the normalized nonreciprocal conductances $\Delta\Tilde{G}(\varphi)\equiv\Delta G(\varphi)/\mathrm{max}(|\Delta G(\varphi)|)$ versus angle $\varphi$ for different spin channels and unconventional pairing symmetries.
    (a) $\Delta\Tilde{G}_{\uparrow_{\hat{n}},\uparrow_{\hat{n}}}$ and $\Delta\Tilde{G}_{\downarrow_{\hat{n}},\uparrow_{\hat{n}}}$  versus $\varphi$ for $s~+~\mathrm{i}d$-wave, consistent with the case in Fig.~1(d-e) of the main text.
(b) $\Delta\Tilde{G}_{\uparrow_{\hat{x}},\uparrow_{\hat{x}}}$, $\Delta\Tilde{G}_{\downarrow_{\hat{x}},\uparrow_{\hat{x}}}$,
and (c) $\Delta\Tilde{G}_{\uparrow_{\hat{y}},\uparrow_{\hat{y}}}$, $\Delta\Tilde{G}_{\downarrow_{\hat{y}},\uparrow_{\hat{y}}}$ versus $\varphi$ for s + helical p-wave, consistent with the case in Fig.~1(f-g) of the main text.
    (d) $\Delta\Tilde{G}_{\uparrow_{\hat{z}},\uparrow_{\hat{z}}}$ and $\Delta\Tilde{G}_{\downarrow_{\hat{z}},\uparrow_{\hat{z}}}$  versus $\varphi$ for s + spinful chiral p-wave with $\boldsymbol{d}=\hat{z}\mathrm{e}^{\mathrm{i}\varphi_{\boldsymbol{k}}}$ and $\psi_{un}=0$.
	Parameters of (a-d) are the same as in Fig.~1 of the main text. 
	\label{figscattnum} }
    \end{figure*}
In this section, we provide detailed numerical simulation results to support the symmetry-based microscopic mechanism and the resulting channel-decomposition theory summarized in Table I and derived in Eqs. (8–10) of the main text.

In Fig.~\ref{figscattnum}(c), we numerically calculate different transport channels for an $s+\mathrm{i}d$ pairing state ($l=2$ for the unconventional component) to verify the microscopic mechanism outlined in the first row of Table~I of the main text.
Both the spin-flip $\Delta G_{\downarrow_{\hat{n}},\uparrow_{\hat{n}}}$ and equal-spin $\Delta G_{\uparrow_{\hat{n}},\uparrow_{\hat{n}}}$ contributions each exhibit an angular dependence proportional to $\mathrm{Im}[\psi_{\rm un}(\varphi)]$ with a $\pi/4$ rotation.
Their combination reconstructs the total nonreciprocal charge conductance $\Delta G_c$, fully consistent with the angular structure shown in Fig.~1(e) of the maintext.

For the mixed $s$–$p$ pairing ($l=1$ for the unconventional component), the spin-resolved channel decomposition is illustrated in Figs.~\ref{figscattnum}(b–d). We first consider the time-reversal ($\mathcal{T}$)-invariant case with $\mathrm{Im}(d_n)=0$, corresponding to an $s+$helical $p$-wave superconductor [Fig.~\ref{figscattnum}(b)]. As predicted by the mechanism summarized in the third row of Table~I in the main text, the spin-flip channels $\Delta G_{\downarrow_{\hat{x}},\uparrow_{\hat{x}}}$ and $\Delta G_{\downarrow_{\hat{y}},\uparrow_{\hat{y}}}$ vanish, whereas the equal-spin channels $\Delta G_{\uparrow_{\hat{x}},\uparrow_{\hat{x}}}$ and $\Delta G_{\uparrow_{\hat{y}},\uparrow_{\hat{y}}}$ remain finite, reflecting the $\mathcal{T}$-invariant nature of the $d_x$ and $d_y$ components in the $s+$helical $p$-wave state. Moreover, $\Delta G_{\uparrow_{\hat{x}},\uparrow_{\hat{x}}}$ [$\Delta G_{\uparrow_{\hat{y}},\uparrow_{\hat{y}}}$] exhibits an angular dependence proportional to $d_x(\varphi)$ [$d_y(\varphi)$], with the two patterns related by a $\pi/2$ rotation. Consequently, the equal-spin channels completely reproduce the total nonreciprocal spin conductance shown in Figs.~1(f,g) of the main text.

Furthermore, we consider the time-reversal-symmetry-breaking case of the \(s\)-\(p\) mixed pairing. As exemplified by the \(s+\) chiral \(p\) pairing state in Fig.~\ref{figscattnum}(d), both the equal-spin contribution \(\Delta G_{\uparrow_{\hat{z}},\uparrow_{\hat{z}}}\) and the spin-flip contribution \(\Delta G_{\downarrow_{\hat{z}},\uparrow_{\hat{z}}}\) exhibit the symmetry-predicted \(\pi/2\)-shifted angular dependences corresponding to \(\mathrm{Re}(d_z)\) and \(\mathrm{Im}(d_z)\), respectively.
Crucially, the independent identification of these two components establishes nonreciprocal spin-dependent conductance as a phase-sensitive probe of the complex spin-triplet order parameter. This behavior agrees with the prediction in the third row of Table~I in the main text.

\section{Numerical Analysis of Robustness under Typical Experimental Conditions}
\begin{figure}[bht]
    \centering
    \includegraphics[width=6.6in]{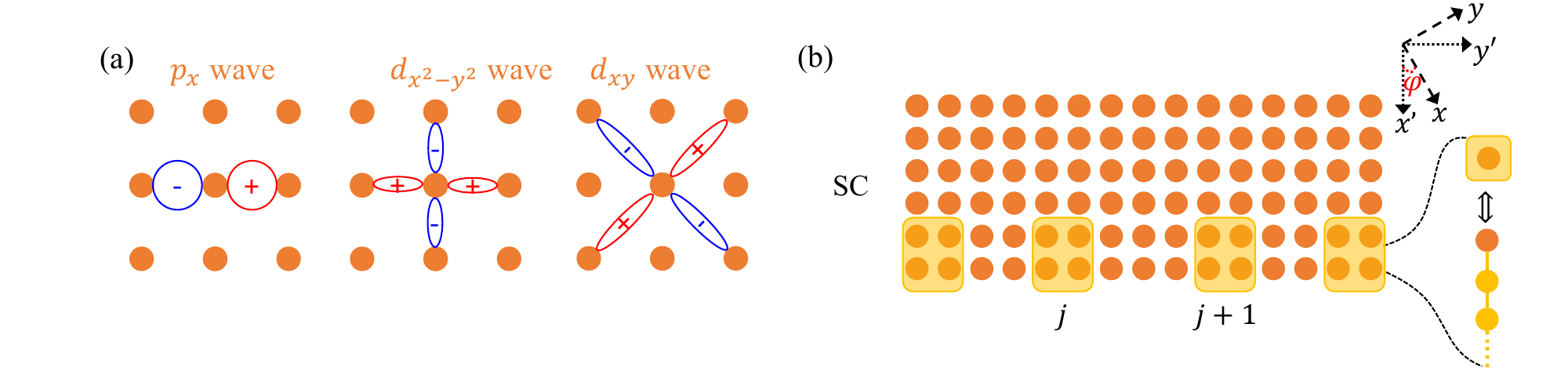}
    \caption{(Color online). (a) Schematic plot of the lattice model to simulate $p_x$, $d_{x^2-y^2}$, $d_{xy}$ wave by difference methods.
    We use the nearest neighbor hopping term to simulate $p_x$ wave, which is similar for $p_y$ wave.
    Meanwhile, we use the nearest neighbor hopping term to simulate $d_{x^2-y^2}$ wave and the next nearest hopping to simulate $d_{xy}$ wave.
    (b) Schematic plot of the lattice model to simulate the nonreciprocal conductance between leads $j$ and $j+1$.
    The superconducting region in the golden rectangle is connected to the lead as the right pattern shows.
	\label{Figlattice} }
    \end{figure}

To compute the nonreciprocal conductance numerically, we employ a finite-difference tight-binding lattice model. 
Under the Andreev approximation, $(E,~|\Delta|)\ll \mu$, one may take 
$\sin(\varphi'_{\mathbf{k}})\approx k_{x'}/k_F$ and $\cos(\varphi'_{\mathbf{k}})\approx k_{y'}/k_F$ 
with $|\mathbf{k}|\approx k_F$, where $k_F$ denotes the Fermi wave vector.
Using the correspondence $\hat{k}_{n}\rightarrow -i\,\partial_{n}$ in the finite-difference scheme, 
we obtain the lattice representations of the $p$-wave and $d$-wave pairing terms shown in 
Fig.~\ref{Figlattice}(a).

In Fig.~\ref{Figlattice}, we present the schematic used to extract the nonreciprocal conductance 
between leads $j$ and $j+1$ in the numerical simulation. 
The axes $x'$ and $y'$ denote the normal and tangential directions at the positions of leads $j$ 
and $j+1$, respectively, while $x$ and $y$ are the corresponding axes for leads $1$ and $2$, 
consistent with Fig.~1(a) of the manuscript. 
The $(x',y')$ coordinate frame is rotated by an angle $\varphi$ with respect to $(x,y)$.
The golden bars represent the metallic leads attached to the superconducting lattice (orange sites).

A detailed illustration of the lead–system coupling is provided in the right part of Fig.~\ref{Figlattice}(b). 
We adopt the wide-band approximation, where the self-energy of lead $j$ is taken as
\begin{eqnarray}\label{eqwideband}
\Sigma_{j,e(h)\sigma} = -\frac{i}{2}\gamma\,\mathrm{I}_{N\times N},
\end{eqnarray}
with $N$ being the number of transverse modes in the lead and 
$\gamma = 2\pi \rho t^2$, where $\rho$ is the density of states in the lead and $t$ is the 
lead–system hopping amplitude.  
Thus, $\gamma$ parametrizes the effective lead–system coupling strength.

A more detailed analysis under typical experimental conditions is necessary.

In this section, we discuss the impact of the lead coupling and the geometric structure of devices, which both are essential since their inevitable influence.
We will show that they do not alter the angular dependence of the nonreciprocal conductance, ensuring the robustness of unconventional pairing detection.
As an illustrative example, we examine the case of nonreciprocal charge conductance in the superconducting system with $s+\mathrm{i}d$ pairing.

{\bf 1. Leads coupling.}
First, we will verify that the lead coupling has no impact on the correspondence between the angular dependence of $\Delta G$ and the angular structure of  unconventional pairing.
Instead, it affects the magnitude of the nonreciprocal conductance\cite{BTK1985}.
Therefore, unconventional pairing can still be effectively identified through nonreciprocal measurements.

\begin{figure}[bht]
    \centering
    \includegraphics[width=6.6in]{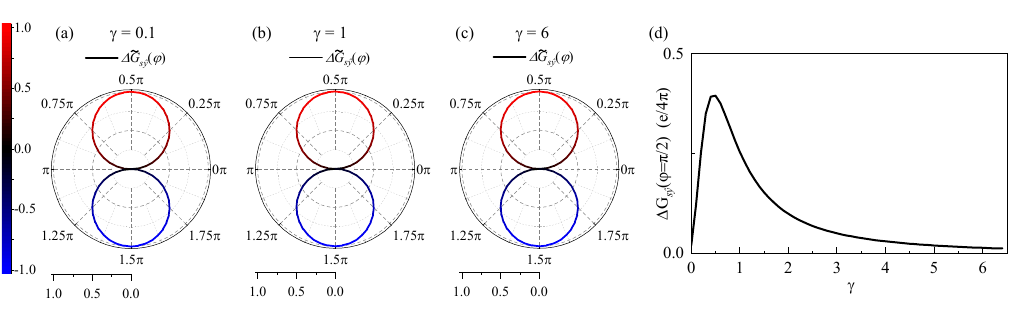}
    \caption{(Color online).
(a–c) Polar plots of the normalized nonreciprocal conductance 
$\Delta \tilde{G}_{s\hat{y}} = G_{s\hat{y}} / \mathrm{max}[|G_{s\hat{y}}(\varphi)|]$ 
as a function of the angle $\varphi$, for (a) $\gamma = 0.1$, (b) $\gamma = 1$, and (c) $\gamma = 6$. 
(d) The magnitude of the nonreciprocal conductance $\Delta G_{s\hat{y}}$ as a function of $\gamma$ at a fixed angle $\varphi = 0$. 
Here we consider an \(s\) + helical \(p\)-wave superconductor with $\boldsymbol{d}=-\hat{x}\mathrm{sin}(\varphi_{\boldsymbol{k}})+\hat{y}\mathrm{cos}(\varphi_{\boldsymbol{k}})$ and $\psi_{un} = 0$. 
The parameters are $B = 1$, $\mu = 1$, $\Delta_s = 0.2$ and $\lambda = 0.02$.
	\label{FigSgamma} }
\end{figure}

According to Eq.\ref{eqwideband}, by tuning $\gamma$, one can control the coupling strength between the leads and the superconducting system.
As shown in Figs.~\ref{FigSgamma}(a--c), we find that the correspondence between the angular dependence of the nonreciprocal conductance and the angular structure of the unconventional pairing persists, with $\gamma$ taken as $0.1$, $1$, and $6$.
On the other hand, the magnitude of the nonreciprocal conductance $|\Delta G|$ varies with the coupling strength, as illustrated in Figs.~\ref{FigSgamma}(d).
In particular, when the coupling between the leads and the superconducting system is optimal, the magnitude attains its maximum value.

{\bf 2. The Geometric Structure.}
We further investigate the impact of device geometry by considering edge roughness, which provides a source of geometric asymmetry. To model this effect, we introduce a random barrier potential at the metal--superconductor interface,
\(U_W(\mathbf{r})\in\left[-\frac{W}{2k_Fa},\frac{W}{2k_Fa}\right],\)
where $k_F$ is the Fermi wave vector, $a$ is the lattice constant, and $W$ characterizes the roughness strength. We consider leads coupled to the edge of the superconducting region.

\begin{figure}[bht]
	\centering
	\includegraphics[width=6.6in]{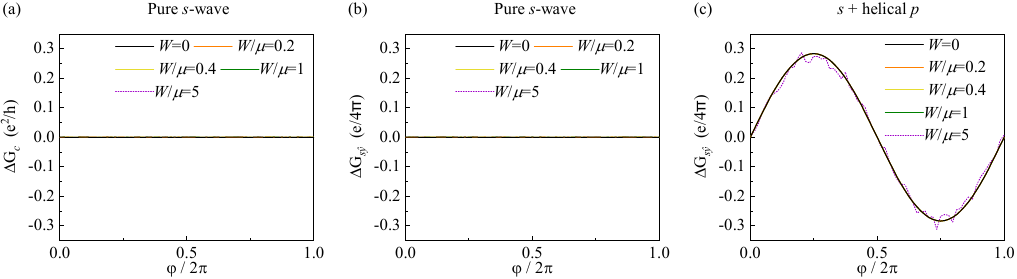}
\caption{
Effect of interface roughness on nonreciprocal transport for leads coupled to the edge of the superconducting region.
(a) Nonreciprocal charge conductance $\Delta G_c(\varphi)$ and
(b) Nonreciprocal spin conductance $\Delta G_{s\hat{y}}(\varphi)$
for a pure $s$-wave superconductor.
The vanishing signals in (a) and (b) demonstrate that geometrical asymmetry induced by interface roughness alone does not generate charge or spin nonreciprocity.
(c) Nonreciprocal spin conductance $\Delta G_{s\hat{y}}(\varphi)$ for an $s+$helical-$p$-wave superconductor characterized by
$\boldsymbol{d}=-\hat{x}\sin(\varphi_{\boldsymbol{k}})
+\hat{y}\cos(\varphi_{\boldsymbol{k}})$ and $\psi_{\mathrm{un}}=0$.
The curves correspond to $W/\mu=0$, $0.2$, $0.4$, $1$, and $5$.
Here, $B=1$, $\mu=1$, $\Delta_s=0.2$, and $\lambda=0.02$.
\label{FigSW} }
\end{figure}

Our symmetry analysis establishes that geometrical asymmetry, including interface roughness, alone cannot generate nonreciprocal transport in a pure $s$-wave state. We verify this prediction numerically by considering a pure $s$-wave state with the same random edge potential. As shown in Fig.~\ref{FigSW}(a) and (b), the charge and spin nonreciprocal conductances, $\Delta G_c$ and $\Delta G_s\hat{n}$, respectively, remain zero for all the roughness strengths considered. This confirms that the nonreciprocal signals discussed below do not originate from geometrical asymmetry or interface roughness alone.

We next consider an $s+$helical-$p$ state and investigate the influence of interface roughness on the nonreciprocal spin conductance $\Delta G_{s\hat{y}}$, as shown in Fig.~\ref{FigSW}(c). For moderate roughness strengths, the effect on $\Delta G_{s\hat{y}}$ is negligible, as indicated by the nearly unchanged curves for $W/\mu=0.2$, $0.4$, and $1$. Notably, for $W<\mu$, the correspondence between the angular dependence of $\Delta G_{s\hat{y}}(\varphi)$ and the angular structure of the unconventional pairing remains intact.
When the roughness strength greatly exceeds the chemical potential, $W\gg\mu$, the random potential induces substantial fluctuations in the interface transparency, resulting in pronounced variations in the angular distribution of $\Delta G_{s\hat{y}}$. Consequently, the correspondence between $\Delta G_{s\hat{y}}(\varphi)$ and the angular structure of the unconventional pairing becomes increasingly indistinct, as illustrated by the curve for $W/\mu=5$ in Fig.~\ref{FigSW}.
Therefore, unconventional pairing can still be effectively identified through nonreciprocal spin transport provided that the interface roughness is not excessively strong, namely, $W<\mu$. Since the nonreciprocity induced by unconventional pairing is a bulk property, moving the leads away from the edge region further suppresses the influence of edge roughness on the nonreciprocal conductance.

\section{Distinguishing Pairing-Induced Nonreciprocity from Other Effects}
The characteristic angular dependence arising from the orbital angular momentum of the hidden unconventional pairing unambiguously distinguishes pairing-induced nonreciprocity from other possible contributions.
In this section, we take the nonreciprocity induced by edge states as a representative example to illustrate how such distinctions can be made.

\begin{figure}[bht]
	\centering
	\includegraphics[width=6.6in]{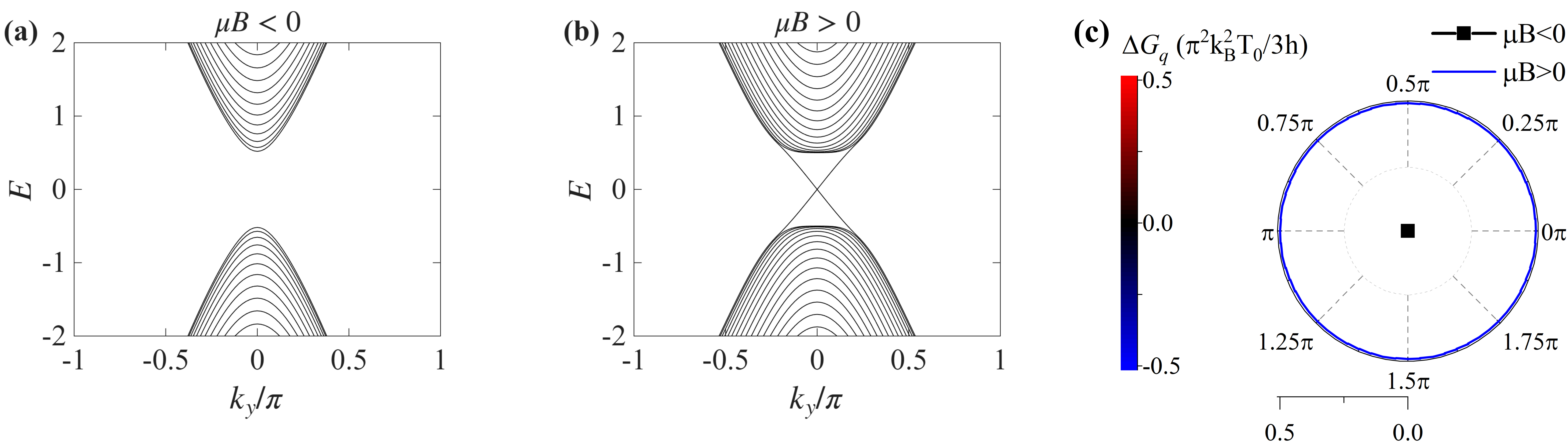}
	\caption{ (Color online).
	(a) Band without edge states for \( \mu = -0.5 \) (\( \mu B < 0 \)). 
	(b) Band with chiral edge states for \( \mu = 0.5 \) (\( \mu B > 0 \)).
	(c) Polar plots of the nonreciprocal thermal conductances $\Delta G_q$ versus angle $\varphi$ for $\mu B<0$ and $\mu B>0$ respectively.
	\label{figtsc} }
\end{figure}

Considering a spinless topological superconductor described by \( H_{SC} = \epsilon_{\boldsymbol{k}}\tau_z + \Delta_{tsc} (k_x \tau_x + k_y \tau_y) \), where the Pauli matrices \( \tau_{x,y,z} \) act in the particle-hole space.
Here we take a quadratic dispersion relation $\epsilon_{\boldsymbol{k}}=\mu-Bk^2$ and $\Delta_{tsc}=\mu/k_F$ with Fermi vector $k_F$.
For $\mu B < 0$, the system is topologically trivial with $\mathcal{N}=0$, and no Majorana edge state exists.
Conversely, for \(\mu B > 0\), the system establishes a topologically nontrivial phase with \(\mathcal{N} = 1\), hosting a Majorana edge state [see Fig.\ref{figtsc}~(a)].
The leads coupled to the edge of the superconductor can then be used to probe the nonreciprocal transport response characteristic of the Majorana edge state [see Fig.\ref{figtsc}~(b)].
As shown in Fig.~\ref{figtsc}(c), the Majorana edge state with  \(\mathcal{N} = 1\) gives rise to a half-quantized nonreciprocal thermal conductance ($G_q=-\frac{\pi^2 k_B^2 T_0}{6h}$) between adjacent leads, independent of the normal angle $\varphi$, which can be identified from the angle-dependent nonreciprocity arising from the unconventional pairing.
This result is consistent with the half-quantized thermal Hall effect induced by the chiral Majorana edge state\cite{Read2000,Sumiyoshi2013,Nomura2012,Kasahara2018,Yokoi2021}.

\begin{figure}[bht]
	\centering
	\includegraphics[width=6.6in]{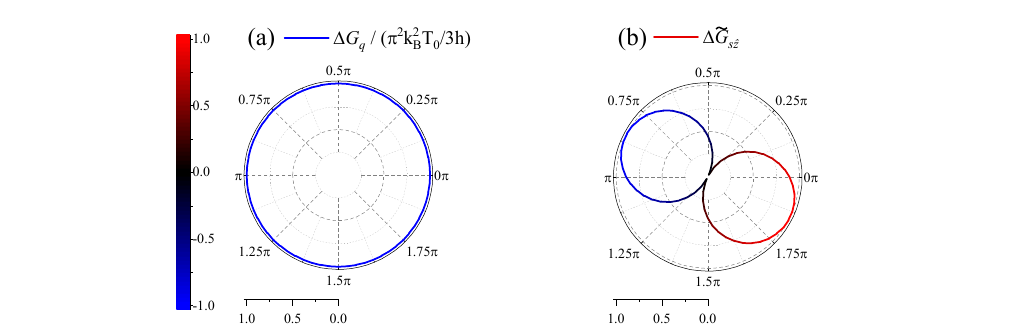}
	\caption{ (Color online).
	(a) Polar plot of the nonreciprocal thermal conductance $\Delta G_q$ versus angle $\varphi$.
	(b) Polar plot of the normalized nonreciprocal spin conductance $\Delta \tilde{G}_{s\hat{z}}=\Delta G_{s\hat{z}}/max(|\Delta G_{s\hat{z}}|)$ versus angle $\varphi$.
	Here we consider an \(s\) + spinful chiral \(p\)-wave superconductor with \(\boldsymbol{d} = \hat{z} \mathrm{e}^{\mathrm{i}\varphi_{\boldsymbol{k}}}\) and $\psi_{un} = 0$. 
    A quadratic dispersion relation $\epsilon({\boldsymbol{k}})=Bk^2$ with $B=1$, $\mu=1$, $\Delta_s=0.2$ and $\lambda=1$ is used for (a-b).
	\label{figtsc2} }
\end{figure}

Furthermore, to illustrate the distinctions between the pairing-induced and edge-state-induced nonreciprocity, we consider a topologically nontrivial $s+$ spinful chiral $p$-wave pairing, which incorporates both types of nonreciprocity.
This can be described by the Hamiltonian in the main text [Eq.~6], with $\lambda > \Delta_s$, $\boldsymbol{d} = \hat{z} \mathrm{e}^{\mathrm{i}\varphi_{\boldsymbol{k}}}$, and $\psi_{\mathrm{un}} = 0$.
This constructs a topologically nontrivial superconducting phase with $\mathcal{N} = 2$, which is reflected in the nonreciprocal thermal conductance $\Delta G_q = -\frac{\pi^2 k_B^2 T_0}{3h}$ induced by the chiral Majorana edge states, independent of the normal angle $\varphi$ [see Fig.~\ref{figtsc2}(a)]. 
Meanwhile, the angular dependence of the nonreciprocal spin conductances reveals the orbital-angular-momentum character of the unconventional spin-triplet pairing [see Fig.~\ref{figtsc2}(b)].
Notably, in such a topologically nontrivial $s+$ spinful chiral $p$-wave superconductor, the unconventional pairing can no longer be treated as a perturbation.
Nevertheless, the correspondence between the angular dependence of the nonreciprocal conductance and the angular structure of the pairing remains valid.

In addition, to observe the nonreciprocity arising from edge states, the leads are coupled to the edge. 
However, as discussed in the previous section, the pairing-induced nonreciprocity is a bulk effect.
As a result, when the leads are coupled to the bulk rather than to the edge, the contribution from edge-state-induced nonreciprocity becomes negligible.
This allows us to focus on the nonreciprocity originating from the pairing.

\bibliographystyle{apsrev4-1}
\bibliography{ref.bib}